\documentclass[preprint,pra,showpacs]{revtex4}
\usepackage{graphicx}
\usepackage{dcolumn}
\usepackage{bm}
\def\wn{cm$^{-1}$}

\begin{document}

\title{Photoassociation of cold atoms with chirped laser pulses:\\time-dependent calculations and  analysis of the adiabatic transfer within a two-state model.}

\author{E Luc-Koenig$^{1}$,  R. Kosloff$^{2}$, F. Masnou-Seeuws$^{1}$ and M. Vatasescu$^{1,3}$}
\affiliation{$^{1}$Laboratoire Aim\'e Cotton, B\^at. 505, 
Campus d'Orsay, 91405 Orsay Cedex, France\\$^{2}$Fritz Haber Research Center for Molecular Dynamics, the Hebrew University of Jerusalem, Jerusalem 91904, Israel\\ $^{3}$Institute of Space Sciences, MG-23, 77125 Bucharest-Magurele, Romania }

\begin{abstract}
This theoretical paper presents numerical calculations for photoassociation of ultracold cesium atoms with a chirped laser pulse and detailed analysis of the results. In contrast with earlier work, the initial state is represented by a stationary continuum wavefunction. In the chosen example, it is shown that an important population transfer is achieved to $\approx 15$ vibrational levels in the vicinity of the v=98 bound level in the external well of the $0_g^-(6s+6p_{3/2})$ potential. Such levels lie in the energy range swept by the instantaneous frequency of the pulse, thus defining a ``photoassociation window''. Levels outside this window may be significantly excited during the pulse, but no population remains there after the pulse. Finally, the population transfer to the last vibrational levels of the ground  $a^3\Sigma_u^+$(6s + 6s) is significant, making stable molecules. The results are interpreted in the framework of a two state model as an adiabatic inversion mechanism, efficient only within the photoassociation window. The large value found for the photoassociation rate suggests promising applications. The present chirp has been designed in view of creating a vibrational wavepacket in the excited state which is focussing at the barrier of the double well potential.
\end{abstract}

\date{\today}

\maketitle

\section{Introduction}

\label{intro}
After ultracold atoms, ultracold molecules are presently a subject of constant interest, stimulated further  by the recent observation of  molecular condensates \cite{jochim03,greiner03,zwierlein03}. This is why the various routes leading to the formation of  ultracold molecules are actively explored  \cite{masnou01}. Starting from molecular beams, two non-optical techniques,  buffer gas cooling of molecules \cite{weinstein98,decarvalho99} and Stark deceleration of polar molecules \cite{bethlem99,bethlem02} now reach temperatures well below 1K. Another route relies on optical techniques: laser fields are used to cool down  alkali atoms, and  to create molecules via the photoassociation reaction \cite{thorsheim87}.  Short-lived  molecules formed in an excited electronic state must then be  stabilized,  by spontaneous emission or other radiative coupling, into bound vibrational levels  of the ground electronic state \cite{fioretti98,takekoshi98,nikolov99,nikolov00,gabbanini00,kerman03}.  The translational temperatures are much lower ($T \le 20 \mu$K) and higher densities are to be expected than with non-optical techniques. An important drawback, however, is that the stable molecules are produced in a superposition of vibrational levels, among which some are very excited. Bringing such molecules to the $v=$0  level of the ground electronic state, thus reaching ultralow vibrational temperatures, is therefore an important issue. More generally, finding new photoassociation and stabilization schemes is an interesting research subject. 

 Up to now, photoassociation experiments have mostly been using continuous lasers: a more active role of the radiation would be to optimize the reaction by  shaping laser pulses.  The aim of the present paper and of the following ones is to explore the feasibility and advantages of experiments using chirped pulses to form  ultracold molecules, therefore bridging the gap between the subjects of ultracold matter and coherent control.

The field of coherent control has developed both theoretically  \cite{kosloff89,zhu98,zhu98b,rice00,shapiro03}  and experimentally \cite{assion98,brixner03,lupulescu04} with quite different applications, such as photodissociation of molecules. It has indeed been demonstrated that shaping laser pulses could enhance considerably the yield of a photodissociation reaction, so that a similar effect is expected for the reverse reaction.

Very few papers consider the photoassociation reaction with  pulsed lasers: in the thermal regime, ultrafast photoassociation  has been studied \cite{marvert95,backhaus99,korolkov96}, showing the validity of the impulsive approximation. In the ultracold regime, several theoretical and experimental papers have proposed a time-dependent study of the photoassociation reaction \cite{mackholm94,boesten96,gensemer98}, eventually considering also  the formation of long-lived ultracold molecules \cite{vardi97}. Most theoretical calculations are using a wavepacket representation for the initial state of the two colliding atoms. In this framework, considering a wavepacket localized at the outer classical turning point of the photoassociated level, our group has discussed the characteristic times for photoassociation of cold cesium atoms \cite{vatasescu01} and the separation of time scales. 

Photoassociation of cold atoms with a chirped-pulsed-laser field was first theoretically  explored  by Vala {\it et al} \cite{vala01}: appreciating the slow nuclear relative motion in ultracold collisions, a chirped pulse was optimized to achieve total transfer of population  under adiabatic following conditions proposed by Cao, Bardeen and Wilson \cite{cao98,cao00}. In these calculations the initial state was represented by a gaussian  wavepacket and a total population transfer was indeed obtained in the region of distances where this wavepacket is localized. The drawback of both this study and  Ref. \cite{vatasescu01} was that they overlooked the precise nature of the initial continuum state. Since the collision energy is so close to threshold, a plane wave description or a superposition of Gaussian wavepackets  do not address the actual shape of the wavefunction. A correct representation of the initial continuum  state of two ultracold colliding atoms should be a stationary continuum state: indeed, experiments with continuous photoassociation lasers identify several nodes of the stationary wavefunction as minima in the photoassociation signal \cite{abraham97,fioretti98}.

The aim of the present paper is to go further in that direction, considering now photoassociation with chirped laser pulses when the initial state is a delocalized continuum wavefunction representing the relative motion of two atoms at very low collision energies $kT$. Such a wavefunction displays very slow oscillations at large distances, and  numerical calculations must then use large spatial grids.  On the other hand, for ultracold processes, where the initial continuum state is very close to threshold, even a very weak light field couples the last bound vibrational levels of the ground electronic surface, which must also be correctly represented on the grid. Then, the numerical calculations need a mapping procedure in order to reduce the number of grid points \cite{slava99,willner04}.

It should be appreciated that the  vibrational levels of the photoassociated molecule, being close to the dissociation limit,  are physically very different from  lower lying vibrational states. Their vibrational periods are orders of magnitude longer and most of the amplitude is concentrated in the outer turning point. As a result, for even  weak light fields, the Rabi period associated with population transfer to the excited state can be shorter than the typical vibrational periods.  For short enough pulses, the relative motion of the two atoms during the pulse duration can be neglected, thus allowing simple interpretations in the framework of the impulsive limit for light-matter interaction \cite{banin94}. This implies that, during the pulse, the wavefunction at each nuclear separation can be decoupled from other positions, so that each distance can be considered as a radiation coupled two-level-system, where transfer of population is analyzed. Another limit now emerging is the limit of adiabatic transfer, where the intensity of the light is sufficient so that the associated Rabi period is smaller than all other relevant timescales.

 The paper shows how a two-state model can be developed to provide analysis of the results within an adiabatic population inversion mechanism,  effective only within a range of energies defining a  photoassociation window. Also discussed is the possibility of avoiding population transfer outside this window.  An example of application is the focussing process, where photoassociation is creating in the excited electronic state a vibrational wavepacket optimized for transfer to a vibrational level of the ground state via a second pulse. 

The paper is organized as follows:  in Section \ref{sec:model} we recall the theoretical model for photoassociation with chirped  laser pulses. In Section \ref{sec:timescales} we discuss the various timescales of the problem. In Section \ref{sec:numres} we give an example of numerical calculations for Cs$_2$ 0$_g^-$ photoassociation, with a chosen chirped pulse, where a few neighbouring  vibrational levels stay populated after the pulse. For interpretation,  section \ref{sec:2state} develops, in the framework of the  impulsive limit,  a two-state adiabatic model  for population inversion within a photoassociation window. In Section \ref{sec:adiab}, the validity of this adiabatic approximation  is discussed.  Section \ref{sec:discuss} is the discussion of the numerical results in the frame of  the simple model for adiabatic population transfer. Finally, in Section  \ref{sec:focus} we show  how the chirp parameter can be chosen in view of focussing a vibrational wavepacket of the photoassociated molecules, such as control of the formation of stable ultracold molecules can be performed. Section \ref{sec:conclu} is the conclusion.

In a forthcoming paper, hereafter referred to as paper II, we shall give more examples for numerical calculations and discuss optimization of the photoassociation yield and of the  cold molecule formation rates.

\newpage   
\section{Theoretical model}
\label{sec:model}
\subsection{The photoassociation reaction}
\label{ssec:reac}
 The photoassociation reaction starts from two cold atoms, at a temperature $T$, colliding in the ground state potential $V_{ground}$, and which absorb a photon red-detuned compared to a resonance line to yield a molecule in a vibrational level of the excited electronic potential $V_{exc}$. We shall consider the example of two cold cesium atoms colliding in the lower triplet  $a^3 \Sigma_u^+(6s+6s)$ potential, and forming a  molecule   in a vibrational level of the  $0_g^-(6s+6p_{3/2})$ potential:
\begin{eqnarray}
Cs(6s^{2}S_{1/2})+Cs(6s^{2}S_{1/2})+\hbar\omega_L
 \rightarrow Cs_{2}(0_g^-(6s^{2}S_{1/2}+6p^{2}P_{3/2};v,J)).
\label{eq:photo}
\end{eqnarray}
The  reaction (\ref{eq:photo}) is the usual representation of the photoassociation, using a continuous laser red-detuned by  $\delta^{at}_{L}$ relative to the $D_2$ resonance line:
\begin{equation}
\hbar\omega_L=\hbar\omega_{at}-\delta^{at}_{L},
\label{eq:deltatL}
\end{equation}
where $\hbar \omega_{at}$ is the energy of the atomic transition 6s $\to$ 6p$_{3/2}$, which excites a single vibrational level of the $0_g^-$ electronic state. In the present work,  we shall treat excitation with a chirped laser pulse characterized by  time-dependent frequency $\omega(t)$ or detuning $\delta^{at}_{L}(t)$, and which populates several vibrational levels.\\

\begin{figure}
{\includegraphics*[width=0.90\textwidth,angle=0]{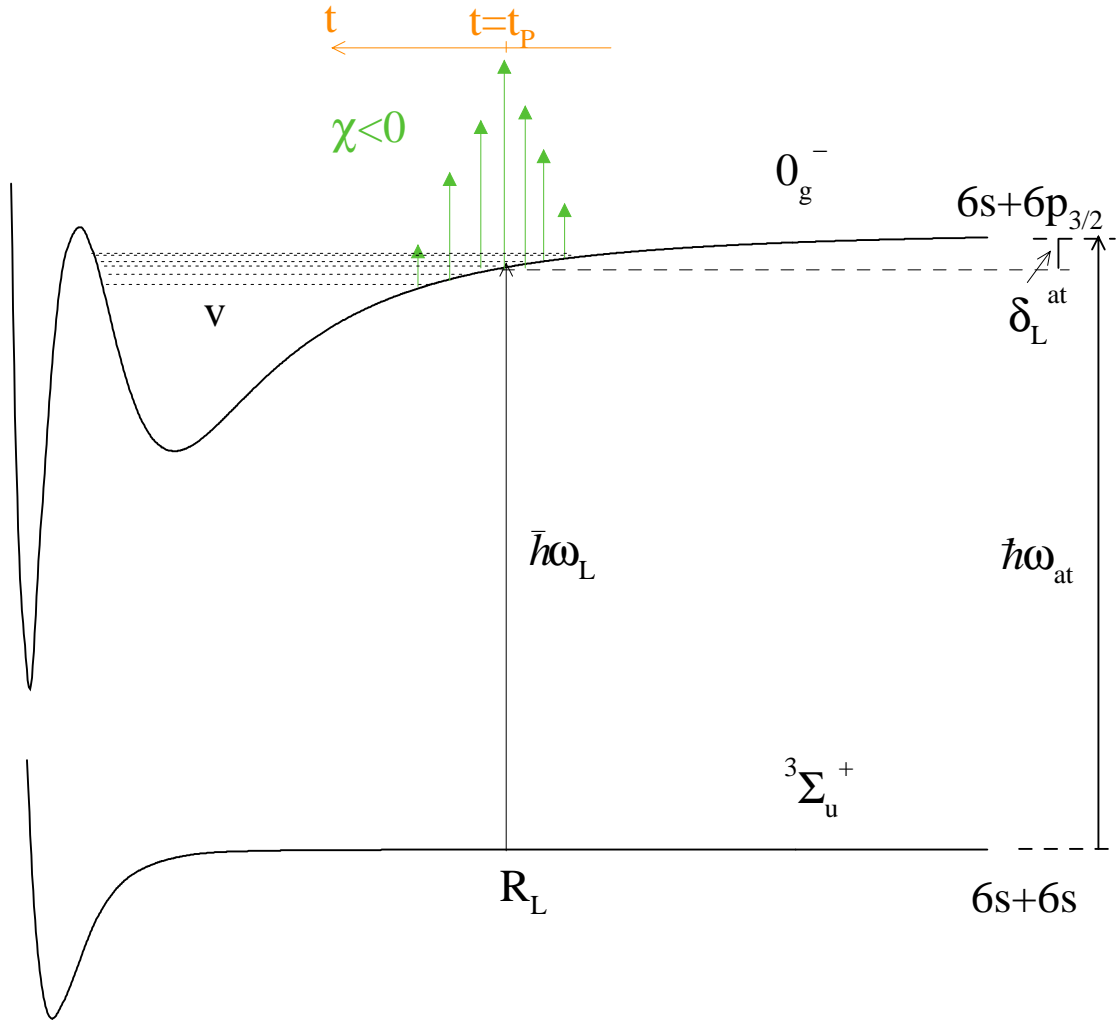}}
\caption{Scheme of the photoassociation process with a negative chirped pulse considered in the present work, illustrated in the case of Cs$_2$. The potentials curves correspond to the ground triplet state a$^3\Sigma_u^+$(6s + 6s) and to 0$_g^-$ (6s + 6p$_{3/2}$) excited electronic state. In the present work, the energy of the initial continuum state is neglected in the definition of the resonance condition. The double well behaviour in the excited curve is a particular feature of the chosen symmetry.}
 \label{fig:pa}
\end{figure}

 As indicated in Fig.\ref{fig:pa}, 
the $0_g^-(6s+6p^{2}P_{3/2})$ potential curve displays a  double well structure with a hump located around 15 $a_0$. The levels populated by the photoassociation reaction correspond to vibrational motion in the outer well: $v_{tot}$ is the vibrational quantum number  in the full potential, while  $v$ is the numbering of the levels in the external well. In the experimental photoassociation spectrum, vibrational levels from $v$=0 to $v$=132 are identified \cite{fioretti99}. For two levels, a tunneling effect is present which has been analyzed from the photoassociation spectrum \cite{mihaela00}.\\

The theoretical model that we are presenting is of course not connected to the particular shape of the excited potential $V_{exc}$.
\subsection{The two-channel coupled equations}
 The dynamics in the ground state and the excited state is described by
the time-dependent Schr\"odinger equation
\begin{equation}
\mbox{$\mathbf{\hat{H}}$} \mbox{$\mathbf{\Psi}(R,t)$} = [\mbox{$\mathbf{\hat{H}_{mol}}$} +
\mbox{$\mathbf{\hat{W}(t)}$}] \mbox{$\mathbf{\Psi}(R,t)$}= i \hbar
\frac{\partial}{\partial t} \mbox{$\mathbf{\Psi}(R,t)$}. 
\label{eq:tse}
\end{equation}
where $\mathbf{\Psi}(R,t)$ is a two-component wavefunction, with ${\Psi}_{ground}(R,t)$ and ${\Psi}_{exc}(R,t)$ describing the radial relative motion of the nuclei in the ground and excited potential respectively.
The molecular Hamiltonian $\mathbf{\hat{H}_{mol}} = \mathbf{\hat{T}} + \mathbf{\hat{V}_{el}}$  is the sum of the kinetic energy operator $\mathbf{\hat{T}}$  and electronic potential energy operator $\mathbf{\hat{V}_{el}}$. The coupling term is written in the dipole approximation:
\begin{equation}
\mbox{$\mathbf{\hat{W}(t)}$}= -\mbox{$\mathbf{\hat{\vec{D}}}(R)$} \cdot \vec{e_L}\ \mbox{${\cal {E}}(t)$}\label{eq:hdip}
\end{equation}
involving the  dipole moment operator  $\mathbf{\hat{\vec{D}}}(R)$ and the electric field defined  by  a polarization vector $\vec{e_L}$ (assumed to be constant) and by an amplitude ${\cal {E}}(t)$. We  define $\vec{D_{ge}}(R)$ from the matrix elements of the dipole moment operator components between the ground and the excited molecular electronic states. Since  the photoassociation reaction occurs at large distances,  we shall neglect the $R-$ dependence,  using the asymptotic value  $\vec{D}_{ge}({R}) \cdot \vec{e_L} \approx D_{ge}^{\vec{e_L}}$ deduced from standard long-range calculations \cite{vatasescu99}. For this reason, the formulas below  are written for a R-independent coupling.

\subsubsection{Excitation by a continuous laser}
For a continuous laser with constant frequency $\omega_{L}/2\pi$, the electric field is 
\begin{equation}
{\cal {E}}(t)= {\cal {E}}_0 \cos(\omega_{L}t)
\end{equation}
 where  ${\cal {E}}_0$ is the amplitude of the field.\\

The explicit temporal dependence of the Hamiltonian $\mathbf{\hat{H}}$ is eliminated in the framework of the rotating wave approximation, defining new radial wavefunctions in both excited and ground state channels through 
\begin{eqnarray}
 {\Psi}^{\omega_L}_{e}(R,t)=\exp(i\frac{\omega_L t}{2})\Psi_{exc}(R,t)\nonumber\\
 {\Psi}^{\omega_L}_{g}(R,t)=\exp(-i\frac{\omega_L t}{2})\Psi_{ground}(R,t),
\label{eqn:rotL}
\end{eqnarray}
and neglecting the high frequency component  in the coupling term. This allows to write the radial coupled equations as
\begin{eqnarray} 
i\hbar\frac{\partial}{\partial t}\left(\begin{array}{c}
 \Psi^{\omega_L}_{e}(R,t)\\
\Psi^{\omega_L}_{g}(R,t)
 \end{array}\right) 
=
 \left(\begin{array}{lc}
 {\bf \hat T} +  V_{e}^{'}(R)  & 
W_L  \\
 W_L & 
 {\bf \hat T} +  V_{g}^{'}(R)
 \end{array} \right) 
 \left( \begin{array}{c}
 \Psi^{\omega_L}_{e}(R,t)\\
\Psi^{\omega_L}_{g}(R,t) 
 \end{array}\right).
 \label{eq:eqcpl1}
 \end{eqnarray}
In  Eqs. (\ref{eq:eqcpl1}), the two non diagonal terms are identical and time-independent,  
\begin{equation}
W_L=\hbar \Omega_L = - \frac {1}{2}{\cal {E}}_0 D_{ge}^{\vec{e_L}}= - \frac {1}{2} \sqrt{\frac{2I}{c \epsilon_0}} 
D_{ge}^{\vec{e_L}}
\label{eq:omega}
\end{equation}
where I is the constant laser intensity, $c$ the velocity of the light and $\epsilon_0$ the vacuum permitivity. 

The diagonal terms in  Eqs. (\ref{eq:eqcpl1}) involve the dressed potentials 
\begin{equation}                                      
 V_{e}^{'}(R)= V_{exc}(R)-\hbar\omega_{L}/2=\bar{V}(R) +\Delta_L(R) ,
 V_{g}^{'}(R)= V_{ground}(R)+\hbar\omega_{L}/2=\bar{V}(R) -\Delta_L(R).
\label{eq:dressed_pot}
\end{equation}
 In Eq. (\ref{eq:dressed_pot}), we have introduced the difference between the two dressed potentials:
\begin{equation}
2 \Delta_L(R)= V_{e}^{'}(R)-  V_{g}^{'}(R)=  V_{exc}(R)- V_{ground}(R)-\hbar\omega_{L},
\label{eq:Delta}
\end{equation}
where  $2 \Delta_L(R) \to \delta_L^{at}$ for large R,  and the mean potential
\begin{equation}
\bar{V}(R)=\frac{V_{exc}(R)+V_{ground}(R)}{2}.
\end{equation}
 When  the dressed-potentials are  crossing at a large distance $R_L$ ($V_{e}^{'}(R_L)=V_{g}^{'}(R_L)$)
\begin{equation}
\Delta_L(R_L)=0,
\label{eq:crossing}
\end{equation}
 the resonance condition for photoassociation into a vibrational level with binding energy $E^L_v$ may be estimated through \cite{masnou01}
\begin{equation}
V_{e}^{'}(R_L)- V_{e}^{'}(\infty) \approx - E^L_v,
\label{eq:reson}
\end{equation}
where we have neglected the kinetic energy of the initial continuum state. Therefore, the outer classical turning point of the photoassociated level should be located close to $R_L$.

 \subsubsection{Chirped pulse and transform-limited pulse}
\label{sssec:chirpdef}

In order to optimize the formation rate of molecules, we shall consider gaussian "chirped" pulses \cite{cao98},  where the laser field ${\cal{E}}(t)$ has a quadratic time-dependent phase $\varphi(t)$, hence  a  time-dependent frequency $\omega(t)$ together with a time-dependent amplitude ${\cal {E}}_0 f(t)$. We have to note that the discussion reported here is general and can be adapted for other types of temporal envelopes than the gaussian one employed here, as for more general chirp rates that could be defined, beyond the linear approximation (Eq.(\ref{eq:freq})) used in the present paper. 
   
\begin{itemize}
  
\item the  field is   
\begin{equation}
{\cal{E}}(t)={\cal {E}}_0 f(t) \cos [\omega_L t+\varphi(t)],
\label{eq:field}
\end{equation}
where  the amplitude involves a gaussian envelope $f(t)$ centered at  $t=t_P$,  
\begin{equation}
f(t)=\sqrt{\frac{\tau_L}{\tau_C}} \exp [-2 \ln 2 (\frac{t-t_P}{\tau_C})^2],
\label{eq:env}
\end{equation}
\begin{equation}
\tau_L=\frac{\tau_C}{\sqrt{1+(\chi\tau_C^2/4\ln 2)^2}} \le \tau_C
 \label{eq:mod}
\end{equation}
with the temporal width $\tau_C$ defined as the full width at  half maximum (FWHM) of the temporal intensity profile ${\cal {E}}_0^2 f(t)^2$. 
 
\item the phase $\varphi(t)$ in Eq. (\ref{eq:field})  has  a second  derivative equal to the linear chirp rate $\chi=\frac{d^2\varphi}{dt^2}$, 
\begin{equation}
\varphi(t)=\frac{1}{2}\chi(t-t_P)^2 -\frac{1}{2} \arccos \frac{\tau_L}{\tau_C}-\omega_Lt_P. 
\end{equation}

\item the frequency of the field, related to the derivative of the phase, varies linearly around the carrier frequency  $\omega_L$:
\begin{equation}
\omega(t)=\omega_L +\frac{d\varphi}{dt}=\omega_L +\chi (t-t_P),
\label{eq:freq}
\end{equation}
 
\end{itemize}

Each pulse is thus defined in the temporal domain by five parameters which are the carrier frequency $\omega_L$, the chirp rate  $\chi$, the temporal center $t_P$, the width $\tau_C$ , and ${\cal{E}}_0$. In the present study of photoassociation, the carrier frequency $\omega_L$ is perfectly defined from an atomic resonance frequency in terms of the detuning $\delta^{at}_{L}$ (see Eq. (\ref{eq:deltatL})). 
 
\begin{figure}
{\includegraphics*[width=0.50\textwidth,angle=0]{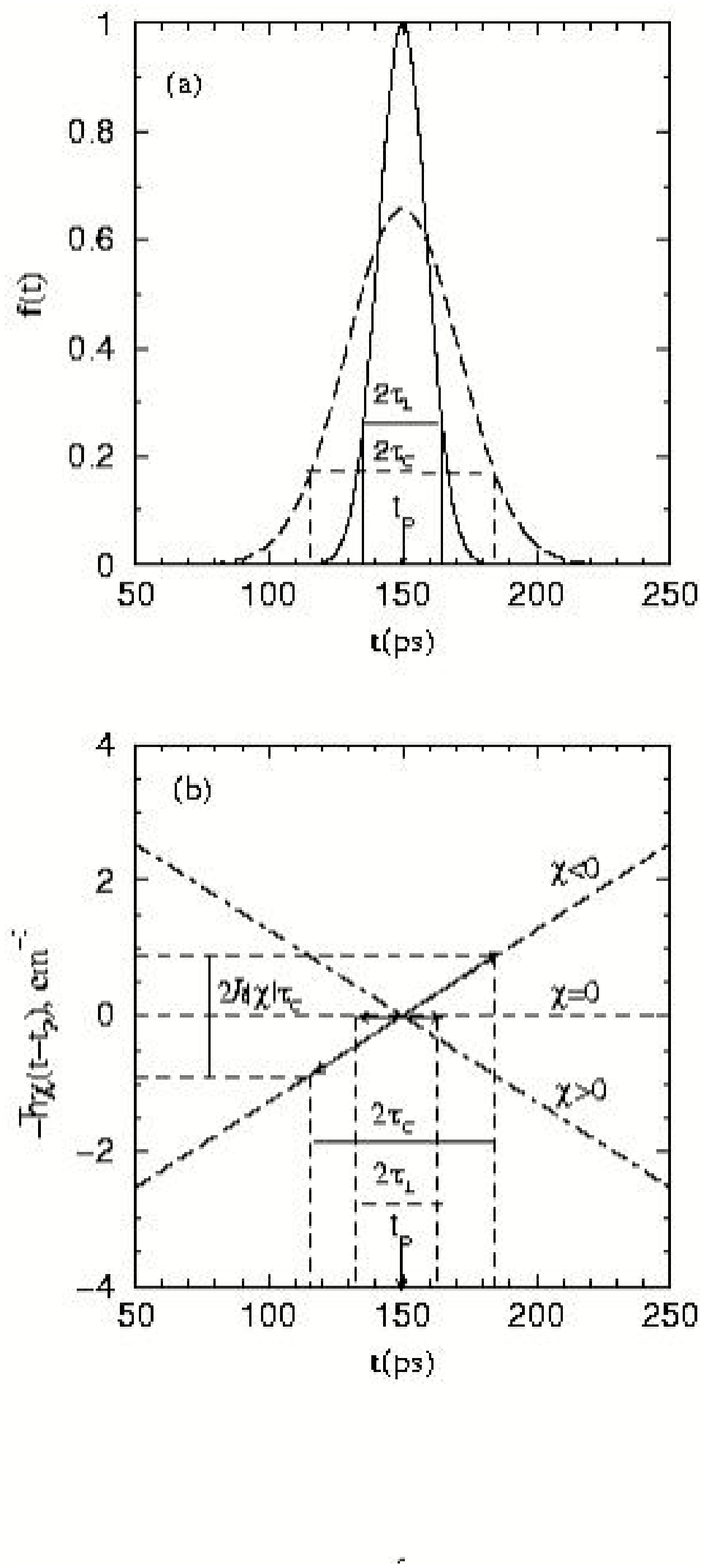}}
\caption{Properties of a chirped pulse. (a) Stretching of the temporal width from $\tau_L$ to $\tau_C$ :  variation of the amplitude for a transform-limited pulse of width $\tau_L$ = 15 ps ( solid line), and for the chirped pulse (broken line) obtained with linear chirp $|\chi|=170$ ps$^{-2}$. Note the reduction of the intensity at maximum, as described in Eq.(\ref{eq:stretch}) in text, and the broadening  to $\tau_C$ = 34.8 ps. The time window [$t_P-\tau_C, t_P+\tau_C$] is indicated by the horizontal line.  (b) Central frequency sweeping:   assuming that the  carrier frequency $\omega_L$ at $t=t_P$ is at resonance with the vibrational level $v=98$ in the Cs$_2$ 0$_g^-$ external well potential, we have indicated the variation of $\omega(t)$ (see Eq.(\ref{eq:freq}) in text) around $\omega_L$ by $\chi (t-t_P)$. }
\label{fig:chirp}
\end{figure}

  Concerning the two last parameters, we must note that for $\chi=0$, $\tau_L=\tau_C$, defining a transform limited  pulse with FWHM of $|f(t)|^2$  equal to $\tau_L$, and   maximum amplitude of the field ${\cal {E}}_0$. Therefore the maximum amplitude of the chirped pulse, ${\cal {E}}_M={\cal {E}}_0 \sqrt{\frac{\tau_L}{\tau_C}}$, is always smaller than ${\cal {E}}_0$: the chirp increases the time-width of the pulse while decreasing its maximum amplitude. The coupling in Eqs.(\ref{eq:eqcpl1}) now becomes time-dependent and reads 
\begin{equation}
W_L f(t) \le W_{max}=W_L \sqrt{\frac{\tau_L}{\tau_C}}=-\sqrt{\frac{\tau_L}{\tau_C}} \frac {1}{2} \sqrt{\frac{2I_L}{c \epsilon_0}} |D_{ge}^{\vec {e_L}}|
\end{equation}
where we have defined a peak intensity $I_L=\frac{c \epsilon_0}{2}{\cal {E}}_0^2$.\\
In the following, we shall assume that the relative phase of the ground and excited electronic wavefunctions is such that $W_L>0$. 
\begin{itemize}
\item 
In contrast, the  chirp does not modify the width of the pulse in the frequency domain, which stays proportional to  $1/\tau_L$. Indeed, $\tilde{\cal E}(\omega)$, which is the Fourier transform of ${\cal {E}}(t)$,  displays a gaussian profile with FWHM $\sqrt{2}\delta \omega$= $4\sqrt{2}\ln 2/\tau_L$, and a phase $\Phi (\omega)$ which is a quadratic function of the frequency, leading to the definition of a linear chirp rate in the frequency domain
 \begin{equation}
 \Phi^{\prime\prime}=\frac{d^2\Phi}{d\omega^2}=\chi\frac{\tau_C^2 \tau_L^2}{(4 \ln 2)^2}.
 \end{equation}
Consequently one has
\begin{equation}
\tau_C^2=\tau_L^2[1+(4\ln2)^2\frac{(\Phi^{\prime\prime})^2}{\tau_L^4}].
\label{eq:stretch}
\end{equation}

\item Finally, the chirp does not change   the energy $E_{pulse}$ carried by the field, which is  proportional to the square of the amplitude ${\cal{E}}_0$ and to the temporal width $\tau_L$ of the transform limited pulse
\begin{equation}
E_{pulse}=\frac{c \epsilon_0}{2}\int_{-\infty}^{+\infty} |{\cal {E}}(t)|^2 dt =\frac{c \epsilon_0}{2} {\cal {E}}_0^2 \tau_L \sqrt{\frac{\pi}{4\ln2}}= \frac{I_L\tau_L }{2} \sqrt\frac{\pi}{\ln 2},
\label{eq:energy} 
\end{equation}
\item This allows us to define a {\it time window}, since for a gaussian pulse 98 $\%$ of the energy is carried during the time interval $[-\tau_C, +\tau_C]$:
\begin{equation}
\frac{c \epsilon_0}{2}\int_{-\tau_C}^{+\tau_C} |{\cal {E}}(t)|^2 dt = 0.98 E_{pulse}
\end{equation}
Since from Eq.(\ref{eq:env}) $f(t_P \pm \tau_C)=\frac{1}{4} \sqrt{\frac{\tau_L}{\tau_C}}$, we may also define a lower limit to the coupling term during the  temporal window and write 
\begin{equation}
\forall (t-t_P) \in [-\tau_C, +\tau_C] \to \frac{1}{4}W_L \sqrt{\frac{\tau_L}{\tau_C}}\le W_L f(t) \le W_L \sqrt{\frac{\tau_L}{\tau_C}}=W_{max}.
\label{eq:bornes}
\end{equation}
The definition of such a time window will reveal very useful for the analysis of the dynamics.
\end{itemize}
We have represented in Fig. \ref{fig:chirp}a  the typical variation of the amplitude of a transform limited pulse and of the  pulse obtained through linear chirp, illustrating both the stretching of the temporal width from $\tau_L$ to $\tau_C$,  the reduction of the peak intensity, and the conservation of the energy carried by the field. The energy range swept during the time-variation of the frequency (see Eq.(\ref{eq:freq})) is illustrated in Fig. \ref{fig:chirp}b.

\subsubsection{Coupled radial equations for excitation with a chirped pulse} 
When considering excitation via a chirped pulse, we may apply the rotating wave approximation as previously at the carrier frequency $\omega_L$. Once the terms  oscillating as $\exp[\pm 2i \omega_L t]$ are eliminated, the resulting coupling term is time-dependent with a phase variation,  so that the coupled equations now are  

\begin{eqnarray} 
&&i\hbar\frac{\partial}{\partial t}\left(\begin{array}{c}
 \Psi^{\omega_L}_{e}(R,t)\\
\Psi^{\omega_L}_{g}(R,t)
 \end{array}\right) \nonumber\\
&&=
 \left(\begin{array}{lc}
 {\bf \hat T} +  \bar{V}(R)+ \Delta_L(R)  & 
W_L f(t)\exp[-i\varphi(t)] \\
 W_L f(t)\exp[i\varphi(t)] & 
 {\bf \hat T} + \bar{V}(R) -\Delta_L(R) 
 \end{array} \right) 
 \left( \begin{array}{c}
 \Psi^{\omega_L}_{e}(R,t)\\
\Psi^{\omega_L}_{g}(R,t) 
 \end{array}\right).
 \label{eq:cplchirp}
 \end{eqnarray}
 However, the existence of a complex coupling term makes the discussion intricated: it is therefore convenient  to modify further the rotating frame by defining new wavefunctions for the excited and the  ground  state,
\begin{eqnarray}
\Psi^{\omega}_{e}(R,t)=\exp(i \varphi/2) \times \Psi^{\omega_L}_{e}(R,t) =\exp(i\frac{\omega_L t + \varphi }{2})\Psi_{exc}(R,t),\nonumber\\
\Psi^{\omega}_{g}(R,t)=\exp(-i \varphi/2) \times \Psi^{\omega_L}_{g}(R,t)=\exp(-i\frac{\omega_L t + \varphi }{2})\Psi_{ground}(R,t),
 \label{eqn:Fcouple}
\end{eqnarray}
thus performing the rotating wave approximation with the instantaneous frequency. The coupled equations now read
\begin{eqnarray} 
&&i\hbar\frac{\partial}{\partial t}\left(\begin{array}{c}
 \Psi^{\omega}_{e}(R,t)\\
\Psi^{\omega}_{g}(R,t)
 \end{array}\right) \nonumber\\
&&=
 \left(\begin{array}{lc}
 {\bf \hat T} + \bar{V}(R)+\Delta_L(R) -\frac{\hbar}{2}\frac{d\varphi}{dt} & 
W_L f(t) \\
 W_L f(t) & 
 {\bf \hat T} + \bar{V}(R)-\Delta_L(R)+\frac{\hbar}{2}\frac{d\varphi}{dt}
 \end{array} \right) 
 \left( \begin{array}{c}
 \Psi^{\omega}_{e}(R,t)\\
\Psi^{\omega}_{g}(R,t) 
 \end{array} \right).
 \label{eq:cplfin}
 \end{eqnarray}
In the following, we shall call $W(t)$ the real time-dependent quantity
\begin{equation}
W(t)=W_L f(t),
\label{eq:W}
\end{equation} 
and $2\Delta(R,t)$ the time-dependent energy difference such that
\begin{equation}
2\Delta (R,t)=2\Delta_L(R)-\hbar \frac{d \varphi}{dt}=2\Delta_L(R)-\hbar \chi(t-t_P)
\label{eq:Del}
\end{equation}
\subsubsection{Resonance window and spectral width}
\label{ssec:reswin}
It is important to note that, in contrast with cw excitation,  the relative position  of the two potential curves at a given distance $R$  now varies linearly as a function of $t$. Hence,  at some distances the upper curve may become the lower or vice versa,  while the position of their crossing point $R_C(t)$, such that
\begin{equation}
 \Delta (R_C(t),t)=0 ; \ \ R_C(t_P)=R_L
\label{eq:reschirp}
\end{equation}
 varies with time around the distance $R_L$ defined in Eq. (\ref{eq:crossing}).\\
 We shall consider situations where during the time window $[t_P-\tau_C , t_P+\tau_C] $,  the crossing distance $R_C(t)$ is spanning a range of distances
\begin{eqnarray}
R_{min} \le R \le R_{max}  
\nonumber \\
R_{min} = R_C (t_P -\frac{\chi}{|\chi|}\tau_C) \ \   \Rightarrow 2 \Delta_L(R_{min})= -\hbar |\chi| \tau_C, \nonumber\\
R_{max} = R_C (t_P + \frac{\chi}{|\chi|}\tau_C) \   \Rightarrow  2 \Delta_L(R_{max})= \hbar |\chi| \tau_C. \label{eqn:window} 
\end{eqnarray}
The definition of such a {\it resonance window} requires two conditions on the width of the pulse and its frequency, to  ensure that the curves keep crossing:
\begin{eqnarray}
  \hbar|\chi| \tau_C \le U^{min}_{eg} \label{eq:prof}\\
\hbar|\chi| \tau_C \le \delta_L^{at}
\label{eq:as}
\end{eqnarray}
where  $-U^{min}_{eg}$ is the minimum value of the difference potential $2\Delta_L(R)$ at $R \le R_L$ , while   $\delta_L^{at}$ the asymptotic splitting.
The  resonance window is defined independently of the pulse intensity. However, it depends upon our choice for the temporal window, and this point will be further discussed below.\\
Writing the resonance condition from Eq.(\ref{eq:reschirp}), which depends upon the central frequency of the pulse,  would be misleading: due to the spectral width,  a range of levels in the neighbourhood of the resonant level can be excited. We shall discuss in more details these two aspects of the excitation with a particular example.  

\section{Time scales}
\label{sec:timescales}
\subsection{Time scales related to the radiation}
\label{ssec:scales}
The two radiative phenomena in our problem are the spontaneous emission time $T_{spont}$ and the Rabi period associated to the coupling with the laser.
We shall  study phenomena at a time scale short compared to $T_{spont}$. 
For  a cw laser, the period $T^L_{Rabi}$ of the ``Rabi oscillations'' associated with the coupling $W_L$  between the two states resonantly coupled is
\begin{equation}
T^{L}_{Rabi}=\frac{\hbar\pi}{W_L}.
\label{eq:rabi}
\end{equation}
Once a pulsed laser is introduced, due to the variation of the peak intensity, this time constant is modified. We define two similar time constants,  at the maximum intensity ($t=t_P$), and at the edges of the time window ($t=t_P \pm \tau_C$), characterizing the dynamics in the vicinity of the instantaneous crossing points $R_L$, $R_{min}$, $R_{max}$:
\begin{eqnarray}
T^C_{Rabi}(t_P)=\frac{\hbar\pi}{W_{max}}=\sqrt{\frac{\tau_C}{\tau_L}}T^{L}_{Rabi} \nonumber\\
T^C_{Rabi}(t_P \pm \tau_C)=\frac{4\hbar\pi}{W_{max}}=4 T^C_{Rabi}(t_P)
\label{eqn:rabiC}
\end{eqnarray}
As a generalization, following Ref.\cite{ashkenazi97}, we shall also use a local time-dependent Rabi period defined by
\begin{equation}
\tau_{Rab}(R,t)=\frac{\hbar\pi}{\sqrt{W^2(t)+\Delta^2(R,t)}}
\label{eq:tauRab}
\end{equation}
 Obviously the pulse duration $\tau_C$ is also an important characteristic of the pulse. For the discussion of the adiabaticity condition, and for the optimization of the pulse, it is interesting to evaluate the time constant associated with the energy range swept during the time window
\begin{equation}
T_{chirp}=\frac{2\pi}{2|\chi| \tau_C}
\label{eq:chirptime}
\end{equation}
This quantity should be compared  to the width of the energy distribution $|\tilde{\cal{E}}(\omega)|^2$, which is $\delta \omega =4 \ln 2 / \tau_L$, and does not depend on the chirp rate. A time scale caracteristic of the spectral width  is then introduced as
\begin{equation}
T_{spect} = \frac{2\pi}{\delta \omega} =  \frac{2\pi}{4 \ln2} \tau_L =2.26 \tau_L .
\label{eq:tspect}
\end{equation} 
The ratio  $\rho$ between those two time constants is bounded, since
\begin{eqnarray}
\rho=\frac{T_{spect}}{T_{chirp}}=\frac{2 |\chi| \tau_C}{\delta \omega}=\frac{2x}{\sqrt{1+x^2}} \label{eq:rho}\\
x= \frac{|\chi| (\tau_C)^2}{4 \ln2}\\
0\le \rho\le 2.
\end{eqnarray}

Therefore the   frequency band swept by the central frequency  due to the chirp is never larger than twice the spectral width. This is an important parameter in the description of the chirp process. 

\subsection{Time constants associated with the dynamics.}

When discussing the vibrational motion of a vibrational level $v$, with binding energy $E_v$ in the excited state, we shall consider the classical vibrational period  estimated from
\begin{equation}
T_{vib}(v)\approx
2 \pi \hbar \frac{\partial v}{\partial E}\approx \frac{4\pi \hbar}{E_{v+1}-E_{v-1}}.
\label{eq:tvib}
\end{equation}
For the discussion on optimization of the chirped pulse, it is also relevant to consider the revival period, defined \cite{averbukh89} as
\begin{equation}
T_{rev}(v)\approx \frac{4 \pi \hbar}{E_{v+1}-2E_v+E_{v-1}}.
\label{eq:trev}
\end{equation}
For the ground state, we shall consider the motion at a low collision energy $E$, described by a scattering length $L$ corresponding  to a time-delay 
\begin{equation}
\tau=\hbar \frac{\partial \delta(E)}{\partial E} \sim \frac{\mu L}{\hbar k},
\label{eq:delay}
\end{equation}
 where we have introduced  the elastic phaseshift $\delta(E)=Arctg(-kL)$, the reduced mass $\mu$ and the wavenumber $k$.

\section{An example of numerical calculations  for photoassociation with a chirped pulse.}
\label{sec:numres}
Since the pulse depends upon many parameters, the way of optimizing them is not straightforward. We give results of numerical calculations in the example already presented in Section \ref{ssec:reac} of photoassociation into several bound levels in the external well of the Cs$_2$ 0$_g^-$ potential, populated by a chirped pulse.
\subsection{The physical problem and its time scales}
\label{ssec:physscales}
 The  photoassociation reaction uses a chirped laser pulse linearly polarized, 
 already illustrated in Fig.\ref{fig:chirp}, for which the parameters are reported in Table \ref{tab:pulsepar}.\\
\begin{table}[h]
\vspace{0.5cm}
\begin{tabular}{|c|c|c|c|c|c|c|c|}
\hline\noalign{\smallskip}
$\delta_L^{at}$ & $I_L$ & $W_L$&$\hbar \delta \omega$& $\tau_L$ & $\tau_C$ &$\hbar \chi$ &$\Phi^{\prime\prime}$\\  
\noalign{\smallskip}\hline\noalign{\smallskip}
2.656 cm$^{-1}$ & 120 kW cm$^{-2}$&0.7396 cm$^{-1}$&0.98 cm$^{-1}$&15 ps&34.8 ps&-0.025 cm$^{-1}$ ps$^{-1}$&-170 ps$^2$\\
 1.21$\times 10^{-5}$ au& &3.37 $\times 10^{-6}$ au&&&&$\chi$=-0.28 $\times$ 10$^{-11}$ au&\\
&&&&&&$\chi$=-4.79$\times$ 10$^{-3}$ ps$^{-2}$&\\
\noalign{\smallskip}\hline
\end{tabular}
\vspace{0.5cm} 
\caption[tab1]
{Parameters for the  pulse excitation considered in the present paper (see section \ref{sssec:chirpdef} and Fig. \ref{fig:chirp}): detuning $\delta_L^{at}$ , intensity $I_L$,  coupling $W_L$, energy associated to the spectral width $\hbar \delta \omega$, temporal widths $\tau_L$ and $\tau_C$, linear chirp parameters $\chi$ and $\Phi^{\prime\prime}$.} 
\label{tab:pulsepar}  
\end{table}

The detuning $\delta_L^{at}$ of the carrier frequency corresponds to  resonant excitation at $t=t_P$ of the level $v$=98 of the  Cs$_2$ 0$_g^-(6s+6p_{3/2})$ potential,  and to a crossing point $R_L=93.7 \ a_0$. Calculations have also been performed by changing the sign of the chirp. For a continuous laser excitation with $\pi$ polarization between the electronic states $^3\Sigma_u^+(6s,6s)$ and 0$_g^-(6s+6p_{3/2})$, and neglecting the $R$-variation of the dipole coupling, the intensity is related to the coupling $W_L$ by Eq. (\ref{eq:omega}), giving $W_L(au)=9.74 \times 10^{-9}\sqrt{I_{W/cm^2}}$ \cite{vatasescu99}.\\

The time scales associated with the radiation are reported  in Table \ref{tab:radtime}. They appear to be of the same order of magnitude: in particular,  $T_{chirp}$ and $T_{spect}$ are similar, since  for the present pulse, the resonance width  $2 |\chi| \tau_c$ is nearly twice the spectral width $\delta \omega$, leading to a value $\rho=1.8$ close to the maximum value which is 2. \\
\begin{table}[h]
\vspace{0.5cm}
\begin{tabular}{|c|c|c|c|c|c|c|c|c|}
\hline\noalign{\smallskip}
$T_{spont}$ & $T^L_{Rabi}$ &$T^C_{Rabi}(t_P)$ &$T^C_{Rabi}(t_P\pm\tau_C)$&$\tau_{Rab}(\infty,t_P)$&$\tau_{Rab}(R=15 a_0,t_P)$&$T_{chirp}$&$T_{spect}$& $\rho$\\  
\noalign{\smallskip}\hline\noalign{\smallskip}
30 ns & 22.5 ps & 34.3 ps &137 ps &12 ps&0.18 ps& 18.9 ps&33.9 ps&1.8 \\
\noalign{\smallskip}\hline
\end{tabular}
\vspace{0.5cm} 
\caption[tab1]
{Time scales associated with the radiative coupling: spontaneous emission time $T_{spont}$, Rabi period $T^L_{Rabi}$,  chirped Rabi periods $T^C_{Rabi}(t_P)$ and $T^C_{Rabi}(t_P \pm \tau_C)$ defined in Eqs.( \ref{eq:rabi}- \ref{eqn:rabiC}). Besides, we have reported the time-dependent local Rabi period $\tau_{Rab}$ at the maximum $t_P$ of the pulse for $R=15 \  a_0$ and $R \to \infty$. Finally, the time constant characteristic of the spectral width of the pulse $T_{spect}$  is reported together with the  chirped  characteristic time $T_{chirp}$ as defined in Eqs. (\ref{eq:chirptime},\ref{eq:tspect}). $\rho$ is the ratio between the resonance width and the spectral width  defined in Eq.(\ref{eq:rho})} 
\label{tab:radtime}  
\end{table}
 The lifetime of the atomic level Cs(6p$^2$P$_{3/2}$) being 30.462 ns, the radiative lifetime of the  photoassociated levels in the external well of the 0$_g^-(6s+6p_{3/2})$ curve is $\approx 30$ ns. In the present calculations  where spontaneous emission is not introduced, we shall  study the evolution of the system at much shorter time scales. Through incertainty relations, it means that the definition of the energy is larger than the natural linewidth of the cesium atom resonance line $\Delta E$= 1.7 10$^{-4}$ cm$^{-1}$= 5.3 MHz. The hyperfine structure splitting of  9193 MHz will not be considered in the present paper, but it has a significant value at the time scale of the problem.\\

At resonance ($\Delta =0$) and for a cw laser, the intensity of the pulse would correspond to a Rabi period of 22.5 ps. However, due to the streching factor $\tau_C/\tau_L$ reducing the maximum intensity, it is increased to 34.3 ps at $t=t_P$. In the wings of the pulse, this time constant is further increased, reaching the value 137 ps at the times $t_P \pm \tau_C$. Such characteristic times should be compared to the times scales  associated to the vibrational motion, and to the collision time.\\  
For the excited potential $V_{exc}(R)$,  the outer well in the 0$_g^-(6s+6p_{3/2})$ curve was fitted to photoassociation spectra  by Amiot {\it et al} \cite{amiot02}. The latter was  obtained \cite{pellegrini03} by matching to {\it ab initio} calculations at short and intermediate range. 
The lower triplet state potential $^3\Sigma_u^+(6s,6s)$ has been chosen in order to ensure a correct value of the scattering length $L$, here taken as $L \approx 525 \ a_0$: the asymptotic behaviour being $C_6/R^6$, where  $C_6$= 6828 au \cite{amiot02},  the short range part extracted from  \cite{spies89} had to be slightly modified \cite{pellegrini03} for that purpose.\\

\begin{figure}
{\includegraphics*[width=0.90\textwidth,angle=0]{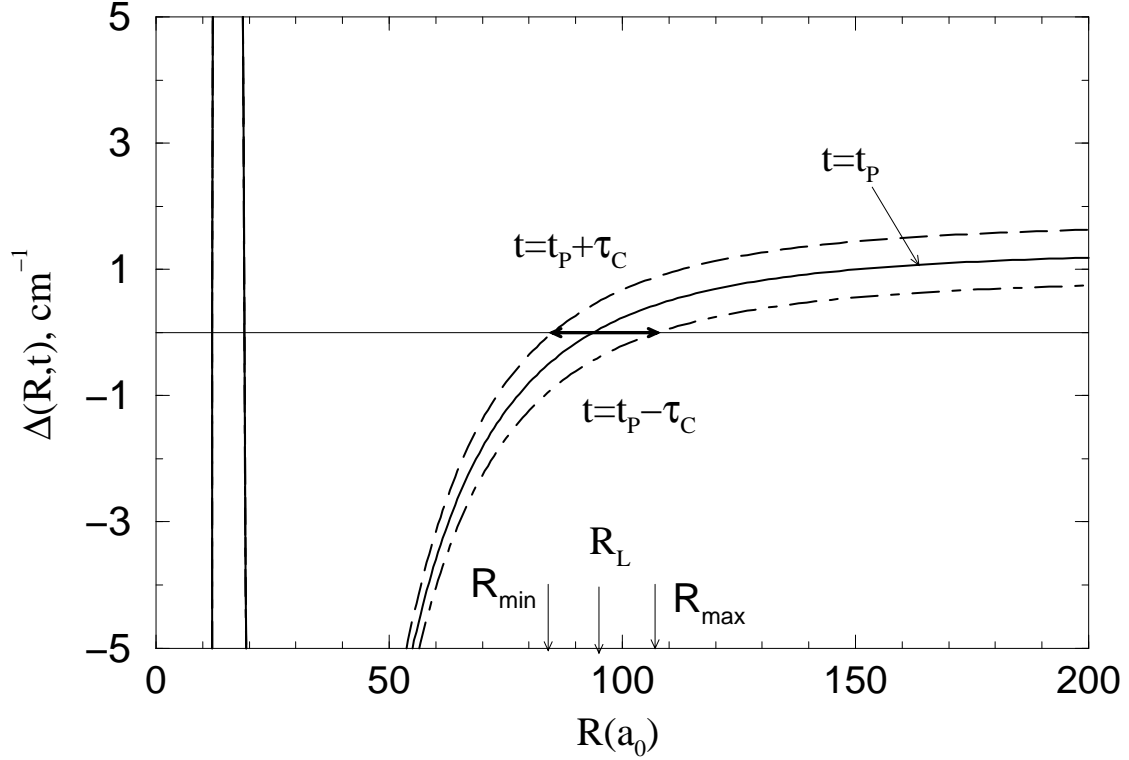}}
\caption{a) Variation of the local time-dependent half energy splitting $\Delta(R,t)$, defined in Eq.(\ref{eq:Del}) as a function of R, for the time values  $t=t_P$ ( solid line),  at the maximum of the pulse;  $t=t_P-\tau_C$, (dash-dotted  line) at the beginning of the time window; $t=t_P+\tau_C$, (broken line) at the end of the time window. The distances where $\Delta(R,t)=0$ define the crossing points $R_L$, $R_C(t_P+\tau_C) \approx R_{min}$ and $R_C(t_P-\tau_C) \approx R_{max}$. b) Photoassociation window: in the reflexion model, the levels $v$=92 to $v$=106  can be populated when the crossing point $R_C(t)$ is sweeping the range [$R_{min}$, $R_{max}$]. The two vertical lines at short $R$ correspond to a local maximum, related to the particular double well structure in the excited curve, having a potential barrier.} 
\label{fig:delta}
\end{figure}

Under the conditions illustrated by the Table \ref{tab:pulsepar}, the time-dependent energy difference $\Delta (R,t)$ between the two dressed potentials, defined in Eq. (\ref{eq:Del}),  is drawn in Fig.\ref{fig:delta}. The long range splitting between the two potential curves is  $2\Delta_L(\infty)=\delta_L^{at}$=2.6 \wn, resulting at $t=t_P$ into a local Rabi period of 12 ps  in the asymptotic region, as reported in Table \ref{tab:radtime}. The energy difference reaches large values at short distances, yielding a very small local Rabi period of 0.18 ps at $R$= 15 a$_0$.  The instantaneous crossing point (defined by $\Delta (R,t)=0$)  varies with time, and we have indicated in  the Figure  the values $R_{min}$=85, $R_L$=93.7  and $R_{max}$=107.4 a$_0$ defined above in Eqs.(\ref{eqn:window}) and discussed in Section \ref{ssec:reswin}. Such distances are close to the outer turning points of the vibrational levels $v$=92, 98 and 106 respectively, for which   Table \ref{tab:excite} displays parameters such as binding energies and  vibrational periods.  The pulse with  negative chirp value described in Table \ref{tab:pulsepar}   has an intantaneous frequency resonant  with the $v$=106 level at $t=t_P-\tau_C$, with  $v$=98 at  $t=t_P$, and $v$=92 at $t=t_P+\tau_C$. Alternatively, when the chirp becomes positive, the resonant condition is verified first by $v$=92 and finally by the upper level $v$=106.\\
 
\begin{table}[h]
\vspace{0.5cm}
\begin{tabular}{|c|c|c|c|c|c|c|}
\hline\noalign{\smallskip}
$v_{tot}$&$v$&$R_{out}$ (a$_0$)&$E_v$ (au)& $E_v$ (cm$^{-1}$) &$T_{vib}$&$T_{rev}$ \\
\noalign{\smallskip}\hline\noalign{\smallskip}  
122&  92 & 85,5 $\approx R_{min}$ &-1.62 $\times$ 10$^{-5}$&-3.57  &196 ps& 10 ns \\
129&98 &   93.7 $\approx R_L$ & - 1.21 $\times$ 10$^{-5}$ & -2.65 &250 ps &15.3 ns\\
137&106&107.48 $\approx R_{max}$&- 0.79 $\times$ 10$^{-5}$ & -1.74  & 350 ps&15.7 ns\\ 
\noalign{\smallskip}\hline
\end{tabular}
\vspace{0.5cm} 
\caption[tab1]
{Characteristic constants for three  levels in the outer well of the 0$_g^-$(P$_{3/2}$) potential curve. $v_{tot}$ is the vibrational number,  while the numbering $v$ is restricted to the levels in the outer well. $-E_v$ is the binding energy, $R_{out}$ is the outer turning point,  $T_{vib}$ and $T_{rev}$ are respectively the classical vibrational period and the revival period as defined in Eqs. (\ref{eq:tvib}-\ref{eq:trev}) in text.   The level $v$=98 is identical to the ($v_0$+2) level considered in Ref. \cite{vatasescu02} on tunneling.} 
\label{tab:excite}  
\end{table}
Due to the linear chirp parameter, an energy range  of about $2\hbar|\chi|\tau_C= 1.74$ cm$^{-1}$  is swept by the laser frequency, corresponding to  15 vibrational levels in the vicinity of v$=98$, for which we have reported data. However, due to the spectral width of the pulse, levels in an energy range of 0.98 \wn on both sides can also be excited, and this will be analyzed with the numerical results.\\

Since the coupling with the ground state may be involving the last least bound levels in the ground state potential curve, we also report in Table \ref{tab:constgr} the binding energies and vibrational periods  for those levels, together with the time-delay for the continuum level at $T=$54.3 $\mu$K, as defined in Eq. (\ref{eq:delay}).    

\begin{table}[h]
\vspace{0.5cm}
\begin{tabular}{|c|c|c|c|}
\hline\noalign{\smallskip}
$v$&$E$ (au)& $E$ (cm$^{-1}$) &$T_{vib}$\\
\noalign{\smallskip}\hline\noalign{\smallskip} 
$v''$=50&-3.87$\times$ 10$^{-7}$& $E$=-.084 \wn & $T_{vib}$=581 ps\\
$v''$=51&-1.23$\times$ 10$^{-7}$& $E$=-0.027 \wn & $T_{vib}$=1460 ps\\
$v''$=52&-1.94$\times$ 10$^{-8}$& $E$=-0.042 \wn & $T_{vib}$={\it 7.8} ns \\
$v''$=53&-2.43$\times$ 10$^{-11}$ & $E$=-5$\times$ 10 $^{-6}$ \wn&\\
continuum &+ 1.71 $\times$ 10$^{-10}$&$E$=+3.77$\times$ 10$^{-5}$ \wn & $\tau  \sim$ 476 ns\\
\noalign{\smallskip}\hline
\end{tabular}  
\vspace{0.5cm} 
\caption[tab2]
{Constants for the last levels in the  $^3\Sigma_u^+$(6s+6s) potentials. The present calculations are not including the hyperfine structure.}
\label{tab:constgr}  
\end{table}

The classical vibrational period has no meaning for the last level, since the wavefunction mainly extends in the classically forbidden region. We should note  the very high value of the time delay in the present problem, due to the large value of the scattering length, demonstrates  the strongly resonant character of the collision, and makes the collision time by far the largest characteristic time in the problem. Looking at Tables \ref{tab:constgr} and \ref{tab:radtime}, we see that the spacing between the level $v''$=53 of the ground state  and the continuum level with  collision energy $E = kT$= 54.3 $\mu$K $\approx$ 3.77 $\times$ 10$^{-5}$ \wn  can be considered as negligible at the scale of the energy uncertainty imposed by  $\approx 30$ ns  radiative lifetime of the excited state.
 
\subsection{Description of the initial state}
  The initial continuum state is represented by a stationary wavefunction in the ground state potential curve and describing the collision of two cold Cs atoms at the energy $kT$ corresponding to  $T$ = 54.3 $\mu$K . It is chosen with a node at the external boundary $L_R$ of the spatial grid (see below). This wavefunction is drawn in Fig. \ref{fig:init}, together with a typical gaussian wavepacket, showing how unlocalized is the initial state in the present work. This is an important modification compared to previous calculations \cite{vala01,vatasescu01} using a gaussian wavepacket:  the population transfer can occur in a very wide range of internuclear distances, well outside the resonance window, and in particular at large internuclear distances where a large density probability is localized. No velocity distribution is considered in the present work, this effect being  treated in paper II.\\

\subsection{Numerical methods}
\label{ssec:num}

Details on the numerical calculations will also be given  in paper II: they involve mapped grid methods \cite{slava99} to represent the radial wavefunctions, using a sine expansion \cite{willner04} rather than the usual Fourier expansion, in order to avoid the occurrence of ghost states. The introduction of adaptive coordinates is necessary to implement a spatial grid with few points (1023), but with a large extension $L_R=19250 \ a_0$ allowing representation of initial continuum states at ultra-low collision energy. Due to the large size of the grid, and the very small kinetic energy of the problem, it is not necessary to put an absorbing boundary condition at the edge of the grid. Since the initial state wavefunction is normalized in a box, the results given below will be depending on the value chosen for $L_R$.\\
The time-dependent Schr\"odinger equation is solved by expanding  the evolution operator $\exp[-i \mathbf{\hat H}t/\hbar]$ in Chebyschev polynomia \cite{kosloff94}. The time propagation is realized by discrete steps with a time increase much shorter  than the characteristic times of the problem, already  discussed in Sec. \ref{ssec:scales} and reported in the Tables \ref{tab:radtime}-\ref{tab:constgr}.  In the present problem, the smallest time scale is the local Rabi period at small distances, $\approx$ 0.18 ps,  which  controls the time step  $\Delta t$.   In typical calculations, we have chosen  $\Delta t \approx 0.05$ ps,  the quantity  $\mathbf{\hat H}\Psi $ being evaluated 112 times at each time step. \\
The dynamics of propagation of the wavepackets in the ground and excited potentials is analyzed by studying the evolution of the population in both surfaces:
\begin{equation}
P_{e}(t)=<\Psi_{e}(R,t) |\Psi_{e}(R,t)>, \ P_{g}(t)=<\Psi_{g}(R,t) |\Psi_{g}(R,t)>
\label{eq:pop}
\end{equation}
A more detailed information is provided by the decomposition of the wavepackets on the unperturbed vibrationals states $v$ of both potential surfaces $S=a^3\Sigma_u^+$ or $0_g^-$:
\begin{equation}
P_{Sv}(t)=|<\Psi_{Sv}(R) |\Psi_{e,g}(R,t)>|^2
\label{eq:popv}
\end{equation}

\subsection{Results}
\label{ssec:res}
All the numerical results presented below are obtained from an initial state wavefunction normalized to unit in a large box of size $L_R$=19250 $a_0$. They correspond to branching ratios of the  photoassociation process towards  different final vibrational levels, eigenstates of the Hamiltonian for the unperturbed molecule. Due to the linear character of the Schr\"odinger equation, the population transferred to a given level is proportional to the normalization factor for the initial state. Therefore, in order to deduce the populations corresponding to an energy-normalized initial continuum state, for $E$= 1.71 $\times$ 10$^{-10}$ au ($T$=54 $\mu$K), the computed populations should be multiplied by the density of states $dn/dE$=$1/(8.87\times 10^{-12}$). Besides, since the calculations concern only one pair of atoms, and are not  velocity averaged,  the values for $P_{e}(t)$   should be scaled  to get an estimation of the population transfer due to a single pulse in a trap containing N atoms: this will be done in Section \ref{sec:discuss}.\\ 
The results of the calculations for the time-dependence of the populations are  presented in Fig. \ref{fig:popu}, both for a negative and a positive linear chirp parameter.  The main conclusions are:
\begin{itemize}
\item There is a  population transfer from the ground $^3\Sigma_u^+$ state to the excited $0_g^-$ state which decreases by  two orders of magnitude from $P_{e}(t_P) \approx$0.032 near the maximum of the pulse, to $P_{e}(t \ge t_P+\tau_C)=3.2\times$ 10$^{-4}$ at  the end of the pulse (see Fig. \ref{fig:popu}a,b). 
\item Calculations considering the same laser pulse with an initial state represented by a gaussian wavepacket normalized to unity yield a much larger value ($\approx 0.59$) for the  final population in the excited state.  However, the ratios between this final population and the probability density in the range 85  $\le R \le$ 110 $a_0$ are similar (0.69 for the gaussian wavepacket, 0.73 for the continuum eigenstate). This result is a signature for the predominant contribution of the resonance window. 
\item The maximum of the transferred population occurs $\approx $ 5 ps before the maximum of the pulse for a negative chirp, and $\approx $ 5 ps after this maximum for a positive chirp  (see Fig. \ref{fig:popu}a,b).
\item After the pulse, most of the population in the ground state is going back to the initial continuum state, but a small fraction ($\approx 3 \times 10^{-4}$) is transferred to the last vibrational levels in the $^3 \Sigma_u^+$ potential, $v''$=51-53 ( Fig. \ref{fig:popu}d). This remaining population is independent of the sign of the chirp.
 The population transfers towards bound levels of the excited $0_g^-$ and of the ground $^3 \Sigma_u^+$ state are comparable.  
\item While many bound levels in the excited  $0_g^-$ potential are excited during the pulse, essentially the levels from $v$= 92 till $v$= 106 remain populated after the pulse (see Fig. \ref{fig:popu}e). They correspond to the range of energy around $v$= 98 swept by the laser during the time window and presented above as  a ``resonance window''. 
\item In contrast,  highly excited levels from $v$=122 to $v=$ 137 are significantly populated during the pulse (see Fig. \ref{fig:popu}e), but the population vanishes at the end of the pulse. 
 \item Whereas the oscillatory pattern during the time window markedly depends upon the sign of the chirp, the final population in the excited state, as well as the vibrational distribution, is nearly independent upon this  sign.
\end{itemize}
Such results demonstrate the existence of a ``{\it photoassociation window}'' including all the levels in the energy range between $v$= 92 and  $v$= 106  where population transfer is taking place. We shall  provide interpretation for this result in the following section.

\section{A two-state model for adiabatic population inversion.}
\label{sec:2state}
\subsection{Impulsive approximation.}
For all the following developments, we shall use the impulsive approximation \cite{banin94}, assuming that the relative motion of the two nuclei is frozen during the pulse duration. The kinetic energy operator being neglected in Eq. (\ref{eq:cplfin}), the two-level Hamiltonian becomes
\begin{eqnarray}
\mathbf {\hat{H}} \sim \mathbf {\hat{H}-\hat{T}}= \mathbf {\hat{H}^i}=
\left(\begin{array}{lc}
 \bar{V}   & 0\\
 0 &  \bar{V} 
 \end{array} \right) +
\left(\begin{array}{lc}
\Delta (R,t)  & W(t)\\
 W(t) &  - \Delta (R,t)
 \end{array} \right)
\label{eq: hamdiab}
\end{eqnarray}
where the first term introduces a $R$-dependent phase while the dynamics is contained in the second term. 
\subsection{The adiabatic basis}
When the impulsive approximation is valid, diagonalization of the hamiltonian $\mathbf {\hat{H}^i}(R,t)$ at each distance $R$ will define a new representation, in the framework of a standard radiation-driven two-level system \cite{ashkenazi97,eberly87}. In Eqs.(\ref{eq: hamdiab}), the notations $W$ and $\Delta$ are similar to the ones used in the textbook by Cohen-Tannoudji {\it et al} \cite{cohen73} to describe a two-state model. Defining a time-dependent local Rabi frequency 
\begin{equation} 
\hbar \Omega(R,t) = \sqrt{\Delta^2(R,t)+W^2(t)},
\label{eq:Omega}
\end{equation}
 diagonalization of  $\mathbf {\hat{H}^i}$ yields two eigenenergies $E_+$ and $E_-$
\begin{equation}
E_{\pm}(R,t)= \bar{V}(R,t) \pm \hbar \Omega(R,t)
\end{equation}
with two eigenfunctions, hereafter referred to as {\it adiabatic},  ${\cal{F}}_+^{\omega}$, ${\cal{F}}_-^{\omega}$. 
 
The set of two {\it adiabatic} functions is deduced from the {\it diabatic} functions  $\Psi^{\omega}_e, \Psi^{\omega}_g$ by a  rotation 
$\mathcal{R}$ \cite{cohen73}:
\begin{eqnarray}
\left(\begin{array}{c}
 \mathcal{F}^{\omega}_{+}\\
\mathcal{F}^{\omega}_{-}
 \end{array}\right) \nonumber= \mathcal{R}
\left(\begin{array}{c}
 \Psi^{\omega}_{e}\\
\Psi^{\omega}_{g}
 \end{array}\right); 
\mathcal{R}=\left(\begin{array}{lc}
\cos \frac{\theta}{2}& \sin \frac{\theta}{2} \\
 -  \sin \frac{\theta}{2} &  \cos \frac{\theta}{2}
 \end{array} \right)
\label{eq:rmatrix}
 \end{eqnarray}
 the angle $\theta$ being defined by the relations:
 \begin{equation}
 \sin [\theta(R,t)] =\frac{|W(t)|}{\hbar\Omega(R,t)}, \ \cos [\theta(R,t)] = \frac{\Delta
(R,t)}{\hbar\Omega(R,t)}, \
 \tan [\theta(R,t)] =\frac{|W(t)|}{\Delta(R,t)}
 \label{eq:theta}
 \end{equation}

 The two-channel wavefunction representing the evolution of the system can then be written in the adiabatic representation as
 \begin{equation}
|{\bar \Psi}(R,t)>= a^+(R,t)|\mathcal{F}^{\omega}_{+}>+a^-(R,t)|\mathcal{F}^{\omega}_{-}> 
\end{equation}
where ${a^+}(R,t)$ and ${a^-}(R,t)$ verify
\begin{equation}
i \hbar \frac{\partial}{\partial t} \left( \begin{array}{c}
 a^+(R,t)\\
a^-(R,t) \end{array} \right) =  \left(\begin{array}{lc}
E_+(R,t)& 0 \\
0 & E_-(R,t) \end{array} \right) 
\left( \begin{array}{c}
 a^+(R,t)\\
a^-(R,t) \end{array} \right)  + \frac{i \hbar}{2} \frac{\partial \theta(R,t) }{\partial t} 
\left(\begin{array}{lc}
0 & 1 \\
 - 1 &  0
 \end{array} \right) \left( \begin{array}{c}
 a^+(R,t)\\
a^-(R,t) \end{array} \right).  
\label{eq:nonadeq}
 \end{equation}
We shall discuss now adiabatic evolution when the second term in the r.h.s. of  Eq.(\ref{eq:nonadeq})  is negligible. The validity of such an hypothesis will be discussed below in Sec.\ref{ssec:adia}. 
 
  \subsection{Condition for  population inversion during the pulse duration.}
\label{ssec:invers}
We assume that the effect of the pulse is negligible outside the time interval $t_P-\tau_F \le t \le t_P+\tau_F$, without assumption on the value of $\tau_F$.
 Considering only the first term in the r.h.s. of Eq.(\ref{eq:nonadeq}),  the solution of  Eq. (\ref{eq:nonadeq}) becomes straighforward :  assuming that before the beginning of the pulse, at $t=t_P-\tau_F$, there is no population in the excited state, the evolution of the diabatic wavefunctions is described by
\begin{eqnarray} 
\left(\begin{array}{lc}
 \Psi^{\omega}_{e}(R,t)  \\  \Psi^{\omega}_{g}(R,t)
 \end{array}\right)\ = 
 \left(\begin{array}{lc}
\cos \frac{\theta(R,t)}{2}& -\sin \frac{\theta(R,t)}{2} \\
  \sin \frac{\theta(R,t)}{2} &  \cos \frac{\theta(R,t)}{2}
 \end{array} \right)\nonumber
 \left(\begin{array}{lc}
e^{-\frac{i}{\hbar} \int_{t_P-\tau_F}^{t} E_+(R,t') dt'} & 0 \\
0 & e^{-\frac{i}{\hbar} \int_{t_P-\tau_F}^{t} E_-(R,t') dt'} \end{array} \right) \\
\times
\left(\begin{array}{lc}
\cos \frac{\theta(R,t_P-\tau_F )}{2}& \sin \frac{\theta(R,t_P-\tau_F)}{2} \\
 - \sin \frac{\theta(R,t_P-\tau_F)}{2} &  \cos \frac{\theta(R,t_P-\tau_F)}{2}
 \end{array} \right)
\left(\begin{array}{c}
0  \\ \Psi^{\omega}_{g}(R,t_P-\tau_F)
 \end{array}\right)
\label{evoltft0}
\end{eqnarray}
Introducing the accumulated angle
\begin{equation}
\alpha(R,t)=\frac{i}{\hbar}\int_{t_P-\tau_F}^{t} E_+(R,t') dt'
\end{equation}
the   two diabatic wavefunctions at time $t$ are
\begin{eqnarray}
 \Psi^{\omega}_{e}(R,t) =\ -[i \sin \frac{\theta_t+\theta_i}{2}\ \sin \alpha(R,t)+\sin \frac{\theta_t-\theta_i}{2} \cos \alpha(R,t)]\ \Psi^{\omega}_{g}(R,t_P-\tau_F),\nonumber
 \\  
\Psi^{\omega}_{g}(R,t)=\  [i \cos \frac{\theta_t+\theta_i}{2}\ \sin \alpha(R,t)+\cos \frac{\theta_t-\theta_i}{2}\ \cos \alpha(R,t)]\ \Psi^{\omega}_{g}(R,t_P-\tau_F),
 \end{eqnarray}
where we have introduced the time-dependent angles 
\begin{eqnarray}
\theta_i =\theta(R,t_P-\tau_F),\\
\theta_t =\theta(R,t). 
\end{eqnarray}

\subsubsection{Pulse of finite duration 2 $\tau_F$}
We are interested in the conditions leading to population transfer, at internuclear distance $R$, from the ground to the excited state, for a pulse starting at $t=t_P-\tau_F$ and stopping at $t=t_P + \tau_F$: 
\begin{equation}
W(t_P-\tau_F)=W(t_P + \tau_F)\approx 0.
\end{equation}
 
From Eqs.(\ref{eq:theta}), we see that the rotation angles are such that
\begin{eqnarray}
\sin \theta(R,t_P-\tau_F)=\sin \theta(R,t_P + \tau_F)=0\\
\cos \theta(R,t_P-\tau_F )=\frac{\Delta(R,t_P-\tau_F )}{|\Delta(R,t_P-\tau_F )|};\ \  \cos \theta(R,t_P + \tau_F )=\frac{\Delta(R,t_P + \tau_F )}{|\Delta(R,t_P + \tau_F )|}
\label{eq:signtheta}
\end{eqnarray}
According to the sign of the time-dependent energy-splitting,  the angles $\theta(t_P-\tau_F )$ and $\theta(t_P + \tau_F)$ can take the values 0 or $\pi$. When the two angles have the same value, the correspondance between diabatic and adiabatic states is the same before and after the pulse, {\it i.e} $|\Psi^{\omega}_{g,e}(R,t_P + \tau_F )| = | \Psi^{\omega}_{g,e}(R,t_P - \tau_F)|$. In contrast,  when the sign of $\Delta(R,t)$ defined in Eq. (\ref{eq:Del})  changes during the pulse, the two states are reversed. This population inversion takes place at a distance $R$ provided that the level splitting between the dressed potentials  verifies
\begin{equation}
 2|\Delta_L(R)| < \hbar |\chi| \tau_F.
\label{eq:adiabinv}
\end{equation}
Then,

\begin{equation}
|\theta(R, t_P+\tau_F) - \theta(R,t_P-\tau_F )|=\pi  \Rightarrow  \ \  |\Psi^{\omega}_{e}(R,t_P+\tau_F ) = | \Psi^{\omega}_{g}(R,t_P-\tau_F )|.
\end{equation}
For a given pulse (fixed values for $\delta_L^{at}$, $\chi$ and $\tau_F$), the condition (\ref{eq:adiabinv}) for population inversion shows that the  transfer will take place in a region of internuclear distances around $R_L$ (such as $\Delta_L(R_L)=0$), the extension of which depends upon the difference $2|\Delta_L(R)|$ between the two potentials and upon the energy range $\hbar |\chi| \tau_F$ swept during the pulse.

In most photoassociation experiments the lasers are tuned so that $R_L$ is located at large distances, where the difference $2|\Delta_L(R)|$ between the ground and excited potential curves is very small, so that we may predict a large photoassociation window. 

\subsubsection{Extension to a gaussian pulse.}
As it was shown in Sec.\ref{sssec:chirpdef}, when the chirped pulse is gaussian, most of the intensity is carried during the time window $(t-t_P) \in [-\tau_C, +\tau_C]$, which suggests  that the population inversion is mainly realized during this temporal window. Therefore, the condition for population inversion during the time window is written introducing  $\tau_C$ instead of $\tau_F$ in the relation (\ref{eq:adiabinv}).  \\
\begin{itemize}
\item For internuclear distances $R$ such that $\Delta (R, t_P-\tau_c) >0$, the angle $\theta_i (R, t_P-\tau_C) \approx 0$, and the diabatic functions are
\begin{eqnarray}
\Psi^{\omega}_{e}(R,t)=-\sin \frac{\theta_t}{2} e^{+i \alpha(R,t)} \Psi^{\omega}_{g}(R,t_P-\tau_C) \nonumber \\
\Psi^{\omega}_{g}(R,t)= \cos\ \frac{\theta_t}{2} e^{+i \alpha(R,t)} \Psi^{\omega}_{g}(R,t_P-\tau_C)
\label{eq:diabneg}
\end{eqnarray} 
\item In contrast, when  $\Delta (R, t_P-\tau_c) < 0$, $\theta_i (R, t_P-\tau_C) \approx \pi$, and 
\begin{eqnarray}
\Psi^{\omega}_{e}(R,t)=\cos \frac{\theta_t}{2} e^{-i \alpha(R,t)} \Psi^{\omega}_{g}(R,t_P-\tau_C) \nonumber \\
\Psi^{\omega}_{g}(R,t)= \sin \frac{\theta_t}{2} e^{-i \alpha(R,t)} \Psi^{\omega}_{g}(R,t_P-\tau_C)
\label{eq:diabpos}
\end{eqnarray} 
\end{itemize}

Let us emphasize that in the adiabatic following of an instantaneous eigenstate of the system, since the angle $\theta_i$ stays strictly equal either to 0 or $\pi$, the population in the ground or in the excited state does not depend upon the accumulated phase $\alpha(R,t)$. This is not valid when $\sin \theta_i \neq 0, \cos \theta_i \neq \pm 1$, since the population at a given $R$ value in the excited channel exhibits Rabi oscillations with a time period corresponding to a $\pi$ variation of the accumulated angle $\alpha(R,t)$. The latter situation requires that the laser field is turned on suddenly.\\

The population inversion condition defines a  range of distances where photoassociation is taking place,  equivalent  to the resonance window defined above in Eqs. (\ref{eqn:window}). However, the validity of such an interpretation relies upon the validity of the adiabatic approximation both inside and outside the window, and of the choice for the temporal window.\\

The conclusions are independent of the sign of the chirp, which is no longer true when non-adiabatic effects are involved.

\section{Adiabaticity regime in the photoassociation window and condition to avoid Rabi cycling at long range }
\label{sec:adiab}
 \subsection{The adiabaticity condition}
\label{ssec:adia}
 From Eq.(\ref{eq:nonadeq}), we see that the system will present an adiabatic evolution provided the non-diagonal term is negligible compared to the energy difference ($E_+-E_-$):
 \begin{equation}
|\frac{\partial \theta }{\partial t}| \ll 4 \Omega(R,t)
\label{adiabcond1}
\end{equation}
 From Eq. (\ref{eq:theta}) we may derive  ${\partial \theta }/{\partial t}=[\Delta (dW/dt) - W ({\partial \Delta}/{\partial t})]/[(\hbar \Omega)^2])$.  From  Eqs. (\ref{eq:W},\ref{eq:Del}), the time-derivatives of the coupling  term $W(t)$ and of the level-splitting $\Delta(R,t)$  are easily derived, leading to the explicit form of the condition (\ref{adiabcond1}),
 \begin{equation}
\hbar |- \frac{4 \ln 2}{\tau_c^2} (t-t_P) \Delta(R,t)  +\frac{\hbar \chi}{2}| \\
\ll
\frac{4[(\Delta(R,t)) ^2 +(W(t))^2]^{3/2} }{ W(t)}=4\frac{[\hbar\Omega(R,t)]^3}{W(t)}, 
 \label{eq:adiabcond2}
 \end{equation}
that we wrote as an inequality between two quantities with the dimension of an energy-square. The quantity in the left-hand side becomes large in the wings of the pulse (when $dW/dt$ is large), far from resonance (when $\Delta(R,t))$ is large) and for a large value of the chirp rate $\chi$. The quantity in the right-hand side is the smallest at the instantaneous crossing point $R_C(t)$ such that $\Delta(R_C,t)=0$, where $\hbar \Omega(R_C,t)=W(t)$.\\
 The condition (\ref{eq:adiabcond2}) takes simple forms if one chooses particular values of R or t:

 \begin{enumerate}
 \item At \underline{$t=t_P$} , when the coupling reaches the maximum $W_{max}=W_L \sqrt{\frac{\tau_L} {\tau_C}}$, one can write an adiabaticity condition function of R:
 \begin{equation}
 \hbar^2 |\chi| \ll \frac{8  [(\Delta_L(R))^2+(W_{L})^2\frac{\tau_L} {\tau_C}]^{3/2}}{W_{L}\sqrt{\tau_L/\tau_C}} 
 \label{adimax}
 \end{equation}
 \item In particular, when both \underline{$t=t_P$ and $\Delta_L(R_L)=0$},  the condition at the "crossing point" $R_L$ for the two dressed potentials, when the coupling reaches its maximum value, reads
 \begin{equation}
 \hbar |\chi| \tau_C\ll 8 W_{L}^2\frac{\tau_L}{\hbar}. 
\label{adiabcond0}
 \end{equation}
\item It is also worhtwhile to consider the very strict adiabaticity condition at the \underline{instantaneous crossing point $R_C(t)$}: 
\begin{equation}
\hbar^2 |\chi| \ll 8(W(t))^2.
\label{adiabcross}
\end{equation}
The condition (\ref{adiabcross}) will not  be verified in the wings of a gaussian pulse. Nevertheless, if we consider the  time window $|t-t_P|<\tau_C$ already defined above, the coupling parameter  has the lower bound $\frac{1}{4} W_{max}$, so that during the \underline{time window} the condition is simply:
\begin{equation}
 \hbar |\chi| \tau_C \ll  W_{L}^2\frac{\tau_L}{2\hbar}
\label{eq:adiabcond3}.
\end{equation}
The validity of such hypothesis will be discussed below.
\end{enumerate}

\subsection{Validity range  of the adiabaticity condition}
 We shall first ensure that, during the temporal window, the adiabaticity condition is indeed valid all across the photoassociation
 window,  defined both by  Eq. (\ref{eqn:window}) and  by the relation (\ref{eq:adiabinv}) in which $\tau_F=\tau_C$. Then we shall check that no population transfer is occuring outside the window, therefore justifying the definition of a window.  
\subsubsection {Adiabaticity condition within the photoassociation window}
From Eq.(\ref{eq:adiabcond2}), we define an adiabaticity parameter, which should be $\ll 1$, as
\begin{equation}
X(R,t)=\frac{X_n}{X_d}=\frac{( \hbar /\tau_C) |- 4 \ln 2 (t-t_P) / (\tau_C) \Delta(R,t)  +\hbar \chi \tau_C /2|}{4[\hbar\Omega(R,t)]^3/W(t)}.
\label{eq:X}
\end{equation} 
During the time window, and within the photoassociation window, the local instantaneous level splitting verifies $|\Delta (R,t)| \le |\hbar \chi \tau_C|$, so that the numerator $X_n$ has an upper limit $X_n \le \hbar^2 |\chi| (4 \ln 2 +1/2)$, reached at the edges of both the time window and the distance window.  Under the same conditions, a lower limit of the denominator can be found from the pulse shape, considering both  the upper and the lower limits on the coupling  during the time window, defined above in Eq. (\ref{eq:bornes}). Therefore, a sufficient condition for adiabaticity within the photoassociation window is
  \begin{equation}
16 \times (\hbar |\chi| \tau_C) \frac{\hbar (4 \ln 2 + 1/2)}{\tau_L}=52.36 \hbar^2 |\chi|  \sqrt{1+\frac {\chi^2 \tau_C^4}{(4 \ln 2)^2}} \ll W_L^2 \Rightarrow X(R,t)\ll 1.
\label{eq:adiabfin}
\end{equation}
The condition (\ref{eq:adiabfin}) means that the laser intensity is sufficient to yield a coupling much larger than the geometric average between the energy range swept by the pulse (2$\hbar |\chi| \tau_C$) and the energy width of the  power spectrum ${|\tilde{\cal{E}}}(\omega)|^2$, which is  ($\hbar 4 \ln 2 /\tau_L$). 
The same condition  may  be written as an upper limit on the energy swept by the chirped pulse through 
\begin{equation}
52.36 \hbar^2 |\chi| \tau_C \ll W_L^2 \tau_L ,
\label{eq:adiabis}
\end{equation}
where in the rhs of  Eq. (\ref{eq:adiabis}) we recognize the quantity $ W_L^2 \tau_L$, proportional to the energy carried by the field. This energy should be sufficient to cause adiabatic transfer.\\
Finally, it is instructive to rewrite the condition (\ref{eq:adiabis}) in terms of the time scales $T^L_{Rabi}$ and $T_{chirp}$, introduced above in Eqs.(\ref{eq:rabi},\ref{eq:chirptime}), as well as the width $\tau_L$ or the time scale $T_{spect}$ associated to the spectral width of the pulse through Eq.(\ref{eq:tspect}):
\begin{equation}
T^L_{Rabi} \ll  0.24\sqrt{ T_{chirp} \tau_L}= 0.16\sqrt{T_{chirp}T_{spect}}.
  \end{equation} 
 Optimizing the adiabaticity condition requires the increase of the intensity of the  pulse
or the reduction either of  the photoassociation window (large value of $T_{chirp}$) or of  the spectral width of the pulse (large value of $\tau_L$ or $T_{spect}$). Since the population transfer will depend upon the width of the photoassociation window, a compromise has to be chosen.
   
\subsubsection {Adiabaticity in the asymptotic region}
\label{sssec:adiabas}
It is often convenient to ensure that no population is transferred outside the photoassociation window, in the region of distances where the two dressed curves never cross during the time interval $(t-t_P) \in [-\tau_C, +\tau_C$]. In particular,  we have considered the long distance region, where $2\Delta_L(R \to \infty)= \delta^{at}_L$,  
$\delta^{at}_L$ being the detuning relative to the atomic transition, as it is defined in Eq. (\ref{eq:photo}).  In this region, considering upper and lower limits of the coupling and of the level splitting, as in the previous section, the validity of the adiabatic approximation is ensured under the condition

\begin{equation}
X_{as}(t)=\frac{\hbar W_{max}}{4\tau_C} [(\frac{\delta_L^{at}}{2} - \frac{\hbar |\chi|\tau_C}{2})^2 +(\frac{W_{max}}{4})^2]^{-3/2} |2 \ln 2 \delta_L^{at} + \frac{\hbar |\chi|\tau_C(4\ln 2 +1) }{2}| \ll 1.
\end{equation}

This condition can be simplified under some circumstances, which depend upon the relative values of $W_{max}$, $\delta_L^{at}$, and $\hbar |\chi| \tau_C$.

\section{Discussion}
\label{sec:discuss}

From the simple model developped above, we shall analyse further the numerical results of Sec. \ref{sec:numres}.
\subsection{Analysis  of the numerical results in the framework of the two-state model}

\begin{figure}
{\includegraphics*[width=0.60\textwidth,angle=0]{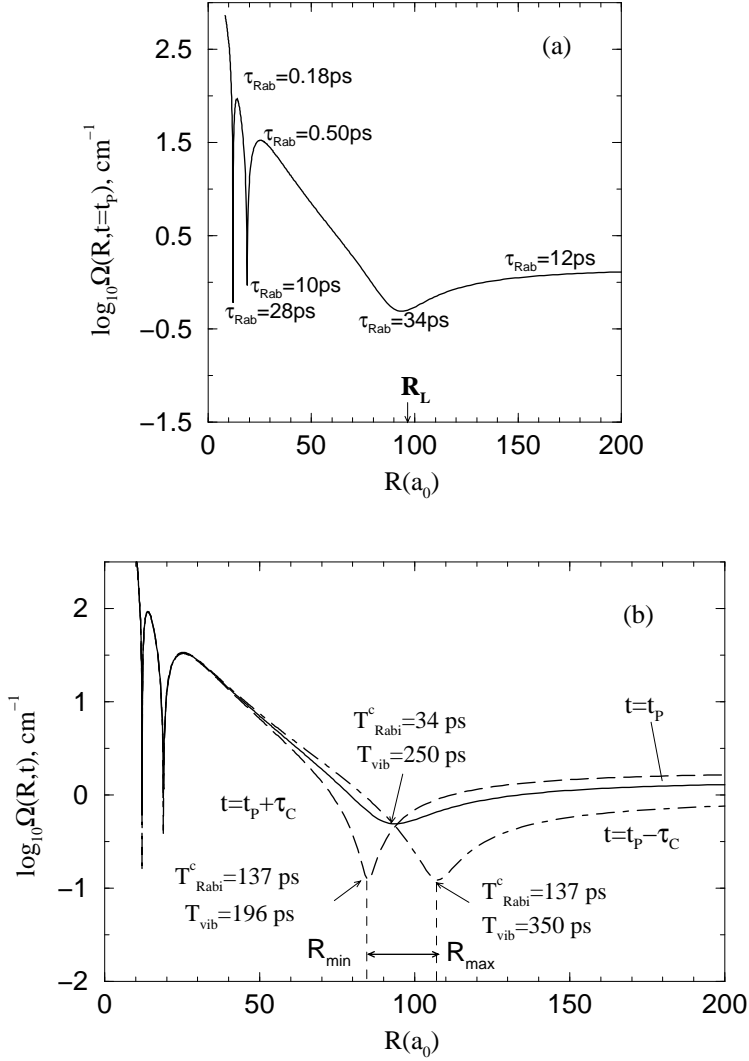}}
\caption{Variation of the local Rabi frequency $\Omega(R,t)$, defined in Eq.(\ref{eq:Omega}) and corresponding to the pulse defined in Table \ref{tab:pulsepar},  as a function of the distance R, for three values of time. (a) Solid line: at $t=t_P$, at the maximum of the pulse. Also indicated are the values of the corresponding local Rabi period $\tau_{Rab}$, see Eq.(\ref{eq:tauRab}) in text. (b) Dash-dotted line: $t=t_P-\tau_C$, at the beginning of the time window. Broken line : $t=t_P + \tau_C$, at the end of the time window. Also indicated are the values of the characteristic times $T_{vib}$ and $T_{Rabi}^C(t_P,t_P \pm \tau_C)$ significant for the vibrational levels resonantly excited at the moments $t_P + \tau_C,t_P, t_P- \tau_C$, which have the outer turning points in $R_{min},R_{L}, R_{max}$.}
\label{fig:omega}
\end{figure}

\begin{figure}
{\includegraphics*[width=0.90\textwidth,angle=0]{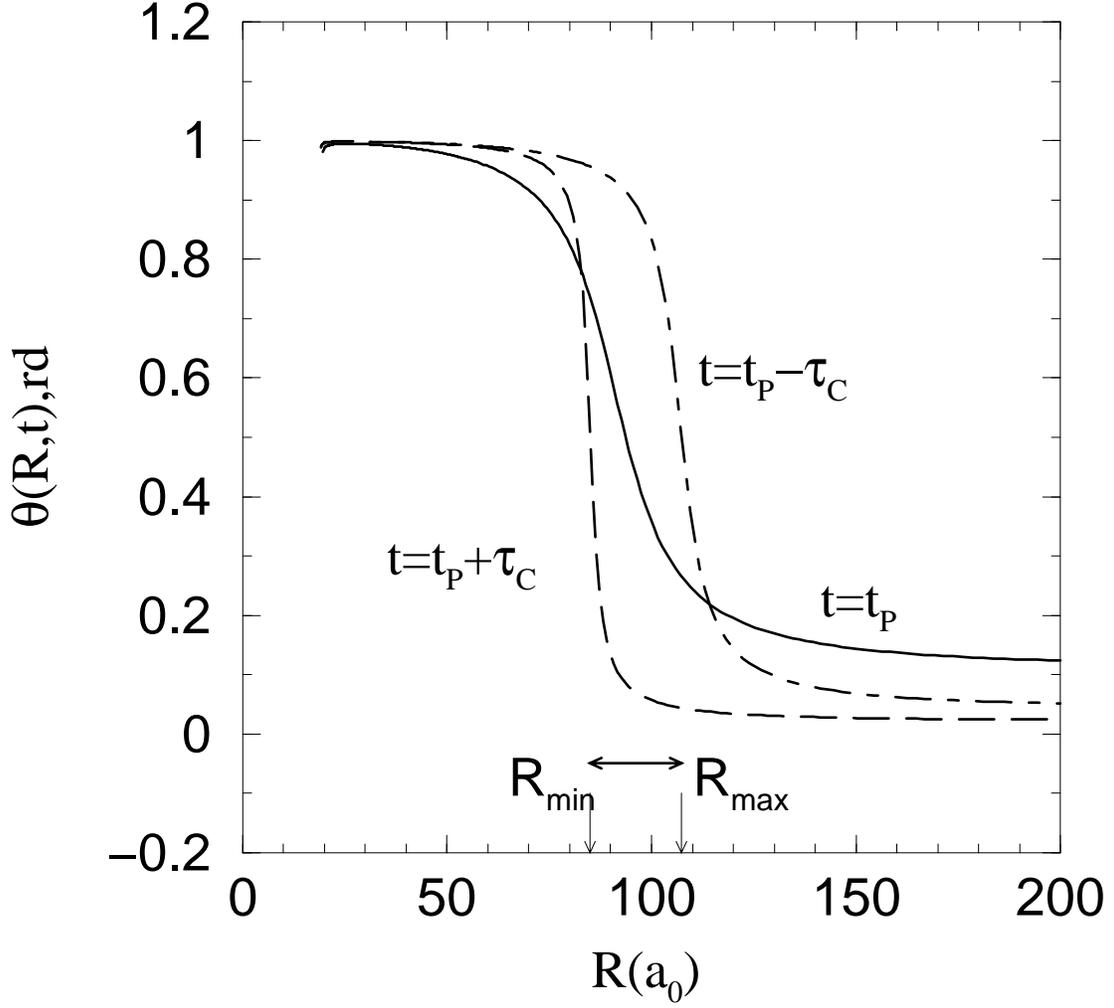}}
\caption{Population inversion and photoassociation window. a) Variation of the angle $\theta(R,t)$, defined in Eq.(\ref{eq:theta}) for the same three time values as in Fig. \ref{fig:omega}. The domain where $\theta(R,t-\tau_C)$ and $\theta(R,t+\tau_C)$ differ by $\pi$ is indicated by an horizontal arrow: it corresponds to population inversion, as discussed in Section \ref{ssec:invers}.}
\label{fig:theta}
\end{figure}

First, we have represented in Fig.\ref{fig:omega} the variation of the local  frequency $\Omega(R,t)$, defined in Eq.(\ref{eq:Omega}), as a function of R for three choices of the time $t=t_P,t_P \pm \tau_C$. The long range minimum  in $\Omega(R,t)$ moves from $R_{max}=$ 110 a$_0$ to $R_{min}$ =85 a$_0$    during the temporal window $[-35,+35]$ ps. Such minima occur at the distances where the local time-dependent energy splitting $\Delta (R,t)$ goes through 0, as illustrated in Fig.\ref{fig:delta}, and therefore define the photoassociation window.  Further minima in the local frequency occur at shorter range, corresponding to inner crossings of the two dressed curves which do not play a role in the photoassociation process, since the amplitude of the initial wavefunction $|\Psi_{init}(R)|$ is negligible at such distances.  Also indicated in  Fig.\ref{fig:omega}a are the values of the local Rabi period at time $t=t_P$, further illustrating the wide range of variation with $R$, from 12 ps at infinity to 34 ps at $R_L$, with very small values (0.18 ps) and larger ones (28 ps) at short range where the two curves part  and cross again. This behaviour is linked to the particular shape of the potential curves, with a double well behaviour for the excited one. Looking at Fig. \ref{fig:omega}b, it is clear that due to the reduction of the pulse intensity in the wings, reducing $W(t)$  by a factor of 4 at $t_P \pm \tau_C$,  the minima in $\Omega(R,t_P \pm \tau_C)$ are deeper at the borders of the time window, multiplying the Rabi frequency by a factor of 4. This is why it will be difficult to verify the adiabaticity condition at the borders of the time window and the photoassociation window. 
   \\
When the adiabatic approximation is valid, the impulsive two-state model described in Sec. \ref{sec:2state}  predicts population inversion within a photoassociation window characterized by a change of $\pi$ between  the angle $\theta(R, t_P-\tau_C)$ and the angle $\theta(R, t_P+\tau_C)$, where $\theta(R,t)$ is defined in Eq. (\ref{eq:theta}). The variation of the angle  $\theta(R,t)$ as a function of $R$ is represented in Fig.\ref{fig:theta} for the same three choices of the time. We see that in a range of distances extending from 85 to 110 a$_0$, $\theta$ varies approximately by $\pi$ from $t_P-\tau_C$ till $t_P+\tau_C$:  a  population inversion can indeed be predicted.\\
 
\begin{figure}
\vspace{2cm}
{\includegraphics*[width=0.70\textwidth,angle=0]{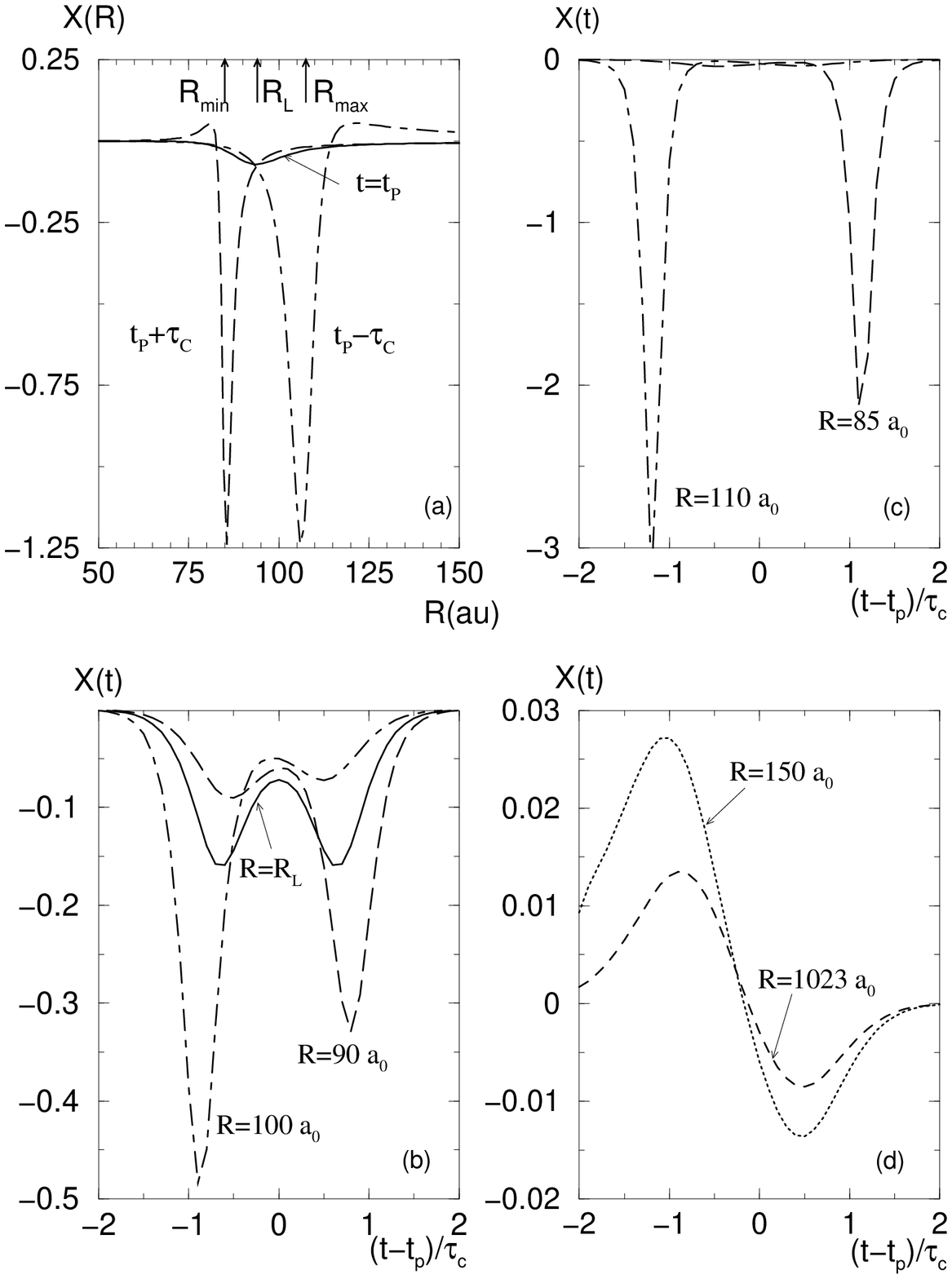}}
\caption{Interpretation of the calculations: validity of the adiabatic approximation within the photoassociation window. The adiabaticity parameter X(R,t), defined in Eq.(\ref{eq:X}) in text, is represented for various values of time and distance.\\ (a)  Variation of $X(R,t)$ as a function of R for $t=t_P$ (solid line), $t=t_P-\tau_C$ (dash-dotted curve), $t=t_P+\tau_C$ (broken curve).\\(b) Adiabatic behaviour within the photoassociation window : Variation of $X(R,t)$ as a function of time for $R=R_L$ (solid curve), $R=90 \ a_0$ (broken curve), $R=100 \ a_0$ (dash-dotted curve). Note that maximum values are reached for $|t-t_P|=\tau_C$.\\
(c) Non-adiabatic behaviour at the border of the window: same as Fig.\ref{fig:adiab}b for $R$=85 $a_0$ (broken curve) and $R$=110 $a_0$ (dash-dotted curve).\\
(d) Adiabatic behaviour in the asymptotic region: same as Fig.\ref{fig:adiab}b for $R$=150 $a_0$ (dotted curve) and $R$=1023 $a_0$ (broken curve).}
\label{fig:adiab}
\end{figure}

The rather good agreement between numerical calculations and the predictions of a simple two-state model seems to indicate that the adiabatic approximation is indeed valid in our problem. In order to get more insight, we have studied the variations of the adiabaticity parameter $X(R,t)$ defined above in Eq.\ref{eq:X}, which are represented  in Fig. \ref{fig:adiab}.  \\
 The adiabaticity parameter takes maximum values at the border of the time window ($t=t_P \pm \tau_C$) and of the photoassociation window $R=R_{min}, R_{max}$, as illustrated in Fig.\ref{fig:adiab}(a). For the pulse with negative chirp described here, we find two adiabatic regions and one non-adiabatic:
\begin{itemize}
\item The adiabaticity condition is verified within  the photoassociation window, as illustrated in Fig. \ref{fig:adiab}(b);
\item It is only approximately  verified at $R \approx R_{min}$ for $t \sim (t_P + \tau_C)$  and at $R \approx R_{max}$ for  $t \sim (t_P - \tau_C)$, where $|X(R$=100 a$_0,t_P - \tau_C)|$  reaches  the value 0.5, as illustrated in Fig.\ref{fig:adiab}(b);
\item It is not verified out of the photoassociation window, in the vicinity of $R_{max}$ and  $R_{min}$, at distances $R$=110 and 85 a$_0$ as illustrated in Fig.\ref{fig:adiab}(c);
\item In contrast, the adiabaticity condition is well verified in the asymptotic region, as illustrated in Fig.\ref{fig:adiab}(d). We should note however that the present pulse has been optimized for that purpose. Other choices yield an important population population transfer at large distances.
\end{itemize} 

The present discussion of the adiabaticity criteria can be attenuated when using a narrower  time window, our choice of $[t_P-\tau_C, t_P +\tau_C]$, corresponding to transfer of as much as 98 $\%$ of the energy in the pulse, being probably too severe. A sufficient choice considers the time window $|t-t_P|\le 0.6 \tau_C$, limited by the two inflexion points in the coupling term $W(t)$, during which 84$\%$ of the intensity is transferred.  Better boundaries for the adiabaticity parameter are then obtained:
\begin{eqnarray}
|t-t_P| \le \tau_C \rightarrow 1.99 \times 10^{-3} < |X| < 29.99 \nonumber \\
|t-t_P| \le 0.6 \tau_C \rightarrow 13.3 \times 10 ^{-3}  < |X| < 1.3849
\end{eqnarray}

\subsection{Discussion}

\begin{figure}
{\includegraphics*[width=0.80\textwidth,angle=0]{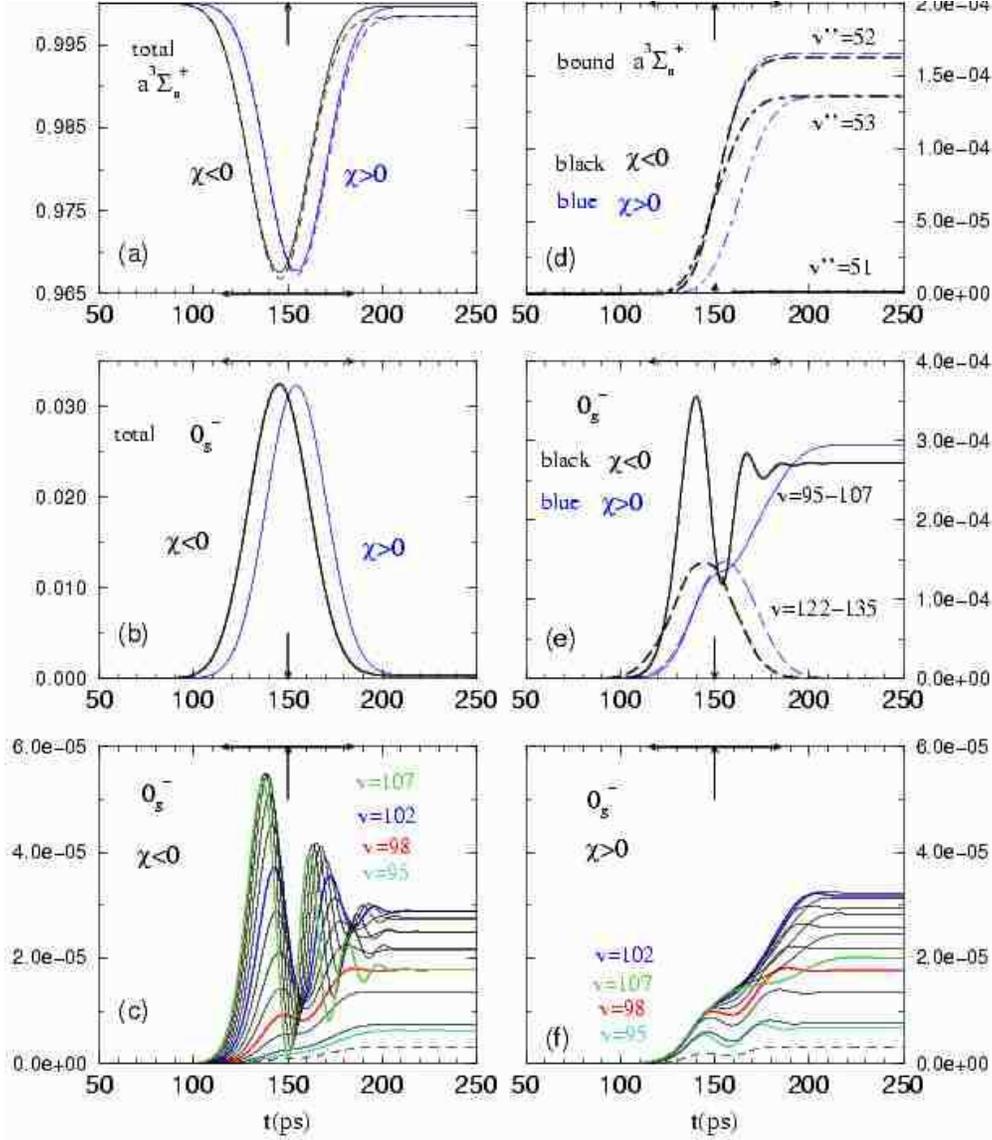}}
\caption{{\footnotesize Time-variation of the computed populations for the ground $^3\Sigma_u^+$ and excited $0_g^-$ electronic states during the pulse duration. The pulse is centered at $t_P=$ 150 ps. Black lines in Fig. (a), (b), (d), (e) correspond to negative chirp, and blue lines to positive chirp.
(a) Full lines: variation of the total population in the  ground $^3\Sigma_u^+$ state. Note that the minimum occurs $\approx$ 5 ps earlier than $t_P=$ 150 ps with  negative chirp, and $\approx$ 5 ps later with positive chirp. Broken lines: variation of the population in the initial continuum level, at $E$=1.72 $\times$ 10$^{-10}$au.
(b) Variation of the total population in the excited $0_g^-$ state for negative  (dark) and positive chirp (blue). (d) Population of the three last bound levels of the $^3\Sigma_u^+$ state: full line $v''$=51, long-dashed line $v''$=52, dash-dotted line  $v''$=53. (e) Variation of the partial sums of populations on the bound vibrational levels in the $0_g^-$ potential: full line for low excited levels, $v$=95 to 107, resonanly excited during the temporal window; dashed line: highly excited levels, $v$= 122 to 135, off-resonance. Black and blue correspond to negative and positive chirp. (c)  Details on the variation of the  individual populations of the $0_g^-$ levels  resonantly  excited  during the temporal window, for negative chirp: in particular red curve, $v$=98, resonant at $t=t_P$, blue $v$=95, dark blue $v$=102, green $v$=107. (f) Same as (c) for a positive chirp.
Populations for an energy-normalized initial state can be deduced by multiplying with 1.12 $\times$10$^{11}$.}}
\label{fig:popu}
\end{figure}

It is instructive at this stage to refine our interpretation of the numerical results presented above in Fig.\ref{fig:popu}. Due to the coupling by the laser, vibrational levels in the $0_g^-$(P$_{3/2}$) are excited. We shall call ``resonant levels'' the $v$=92-106 levels, corresponding to the range of energy swept by the instantaneous frequency $\omega (t)$ during the time window, and ``off-resonance levels'' the other ones, corresponding to energies swept by the wings of the spectral distribution during the time window, or by the whole spectral distribution outside of the time window. Among them, excited levels are numerous, such as the $v$=122-137 range, due to the high density of levels close to the dissociation limit, and to the good Franck-Condon overlap with the initial state. \\
When the chirp is negative, the instantaneous crossing point is moving from $R_{max}$ to $R_{min}$, so that the most excited levels are populated first. This is illustrated for the ``resonant levels'' in Fig.\ref{fig:popu}(c). In contrast, for a positive chirp, since the resonance window is swept from $R_{min}$ to $R_{max}$, the lower levels are populated first, as illustrated for the resonant levels in Fig.\ref{fig:popu}(f).\\

 For  the ``off resonance'' levels in the excited state, a maximum in the population $P_v(t)$ (see Eq. \ref{eq:popv}) appears, independently of the level, 4.45 ps before $t_P$ for $\chi <0$, and 4.45 ps after $t_P$ for $\chi > 0$ (Fig.\ref{fig:popu}(e)). A quantitative evaluation of this advance (resp. delay) can be obtained from analysis of the adiabatic population transfer at large distance $R \ge $ 150 a$_0$, where $2 \Delta_L (R)\approx \delta_L^{at}$, the adiabatic parameter being such that $|X(R,t)|<$ 0.03 (Fig.\ref{fig:adiab}). The diabatic wavefunction in the excited potential $0_g^-$ is given by Eq.(\ref{eq:diabneg}) where both the angle $\theta (R,t)$ and the accumulated angle $\alpha (R,t)$ are independent of $R$. Furthermore the vibrational wavefunction in the excited potential curve is strongly localized in the asymptotic region. Therefore the time-dependence of the population $P_e(t)$ as defined in Eq.(\ref{eq:pop}) scales as $\sin^2   \frac{\theta}{2}(\infty,t)$, reaching a maximum for a time $t_m$ satisfying $\frac{d\theta}{dt}(\infty, t_m)$=0, and corresponding to a zero value of the adiabaticity parameter. Trivial calculations yield the condition
\begin{equation}
t_m-t_P \sim \frac{\hbar \chi \tau_C^2}{4 \ln 2 \delta_L^{at}}
\end{equation}    
 which gives  -4.37 ps for the present pulse  with negative chirp, and explains the symmetry when changing to a positive chirp. The excellent agreement between this estimation and the $\pm$4.45 ps value found in numerical calculations is a signature of the validity both of the adiabatic approximation and of the impulsive approximation in the asymptotic region. Indeed, the vibrational period for levels $v >$ 122 is $T_{vib}(v) \ge$ 780 ps. The maximum is also visible in the variation of total population in Fig.\ref{fig:popu}(b), therefore dominated by asymptotic excitation. \\ 
 
  For the ``resonant'' levels, oscillations in $P_v(t)$ appear only for the levels which come to resonance with the instantaneous frequency $\omega(t)$ before the maximum of the pulse,  at a time $t_C(v)$ such that  $t_C(v) < t_P$. For $\chi <$ 0,  oscillations are therefore observed for the highest levels,  like $v$=107 and 102 in Fig.\ref{fig:popu}(c). For $\chi >$ 0, the population of the lowest levels, like $v$=95 in Fig.\ref{fig:popu}(f) is oscillating. For the $v$=98 level, resonant at the maximum of the pulse, the time-dependence of the population does not depend upon the sign of the chirp (red curve in Fig.\ref{fig:popu}(c,f)), and oscillates weakly. Such oscillations are a signature of significant non-adiabatic effects in the population transfer at the instantaneous crossing point around $t=t_P - \tau_C$. We have shown in Sec.\ref{sec:adiab} how the adiabaticity criterion is most difficult to verify at the instantaneous crossing point and in the wings of the pulse. Away from the adiabatic regime, the passage through resonance does not lead to a total population transfer:  coherent excitation  of a two-level system with linearly chirped pulses results in ``coherent transients'' previously studied in the perturbative limit  \cite{eberly87,zamith01,degert02}. Such transients are governed by interferences between the population amplitude transferred at resonance $t=t_C(v)$, and population amplitude transferred after, for $t>t_C(v)$. The latter amplitude is significant provided the maximum of the pulse occurs after the resonance, so that $t_P>t_C(v)$, resulting into the observed oscillations. The values of the observed periods can correspond to a Rabi period at the instantaneous crossing point:  for $v$=98, resonant at $t=t_P$, the period of 36 ps is close to $\tau_{Rab}(R_L,t_P)$= 34 ps; for the level $v$=102 with a negative chirp, or $v$=95 with a positive chirp, the 25 ps period is associated to a $\pi$ variation of the accumulated angle $\alpha(R,t)$ calculated at R distances corresponding to the outer turning points of these vibrational levels.\\

 Looking back to the derivation of the equations describing adiabatic transfer in Sec.\ref{sec:2state}, we see that in the case when $\sin \theta_i \ne 0 $ ( or $\pi$), i.e. when the adiabatic evolution starts only once the laser has been turned on for a while, the evolution of the population in the excited state is no longer described by Eq. (\ref{eq:diabneg})   (or Eq.(\ref{eq:diabpos})), but verifies
\begin{equation} 
|\Psi^{\omega}_{e}(R,t)|^2=[\sin^2 \frac{\theta_t}{2} \cos^2 \frac{\theta_i}{2}  +   \cos^2 \frac{\theta_t}{2}\sin^2\frac{\theta_i}{2} - \frac{1}{2}  \sin\theta_t \sin\theta_i \cos [2\alpha(R,t)] |\Psi^{\omega}_{g}(R,t_P-\tau_{ad}|^2 
\end{equation}
for $\chi < 0$, where we have called $t_P-\tau_{ad}>t_P-\tau_C$ the time where the adiabatic evolution starts (a similar equation can be written for $\chi > 0$, with change of $\sin$ into $\cos$). \\ \\
The total population at a given $R$ value in the excited channel therefore exhibits Rabi oscillations, with a time period corresponding to a $\pi$ variation of the accumulated angle $\alpha(R,t)$. The modulation rates appearing in Fig.\ref{fig:popu}(c) for the high resonant vibrational levels with large vibrational periods $T_{vib} \sim 350$ ps, which are the first populated with a negative chirp, are much larger than those observed  in Fig.\ref{fig:popu}(f) for the low resonant  vibrational levels, with $T_{vib}\sim 200$ ps, the first populated with a positive chirp. Rabi oscillations being manifested in cases where the vibrational period is much larger than the Rabi period, a frailty in the impulsive approximation can be supposed for levels for which these two characteristic times become comparable. Besides, as discussed by  Banin {\it et al} (1994) \cite{banin94}, the wavepacket created in the excited state being accelerated towards short internuclear distances, the breakdown of the impulsive approximation is more severe for $\chi <$  0 than for  $\chi >$  0. Indeed, for $\chi <$  0, the instantaneous frequency of the laser remains resonant during the motion of the wavepacket and then recycles the population back to ground state \cite{cao00}.\\  
Therefore, the simple model for two-state adiabatic population inversion, developped in the impulsive limit, allows a qualitative interpretation of the numerical results. In particular, the vibrational levels in the excited state where population remains transferred after the pulse are well predicted by this model. Also is the weak dependence of the final population as a function of the sign of the chirp. However, the phase  of the probability amplitude on each level is not independent of the sign of the chirp, so that after the pulse the vibrational wavepacket  evolves in a very different way for positive and for negative chirp. Furthermore, detailed analysis of the numerical results indicates the limits of both the impulsive model and the adiabatic population transfer.

\section {Application:  controlling the formation of ultracold molecules via focussing and improving present experimental schemes}
\label{sec:focus}
\subsection {Focussed vibrational wavepacket}
\begin{figure}
{\includegraphics*[width=0.98\textwidth,angle=0]{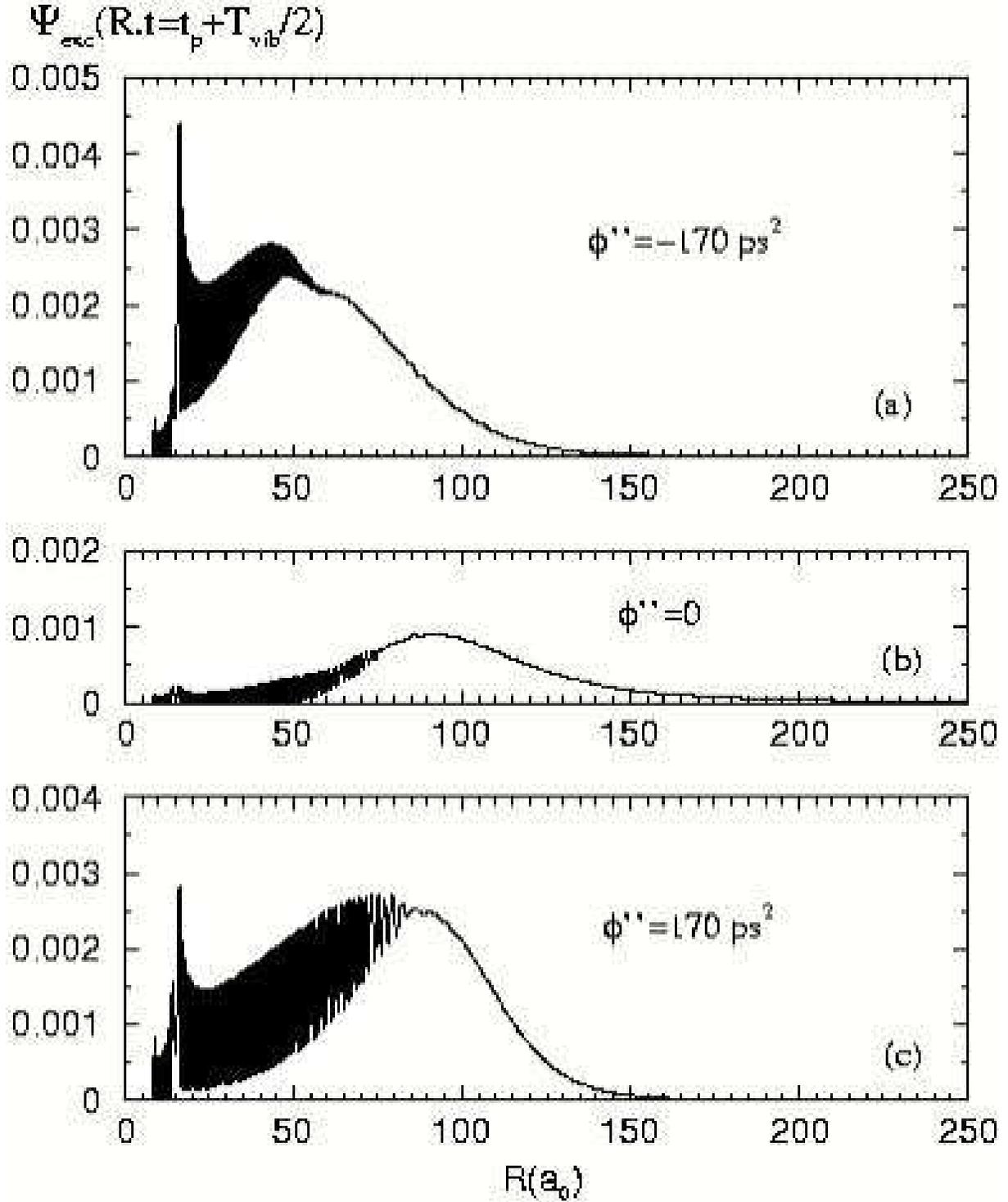}}
\caption{Optimization of the chirp in view of focussing. Variation of the amplitude of the wavefunction in the excited state $|\Psi_{exc}(R, t=t_P+125)|$ , at a delay $T_{vib}/2$ = 125 ps after the maximum of the pulse (i.e. well out of the time window), where $T_{vib}$ is the classical vibrational period for the resonant level $v$=98. (a) negative chirp, for the pulse defined  in Table \ref{tab:pulsepar}; (b) same with $\chi$=0; (c) same with positive chirp parameter.}   
\label{fig:focus}
\end{figure}

The pulse described in Table \ref{tab:pulsepar}, with negative linear chirp parameter $\chi$,  has been designed in order to achieve focussing of the vibrational wavepacket at a time $t_P + T_{vib}/2$, where $T_{vib}/2$= 125 ps is   half the vibrational period of the level $v$=98. The levels in the range from approximately $v$=106 till $v$=92 are populated in turn, the chirp parameter necessary to compensate the dispersion in the vibrational period of the wavepacket being chosen as $\chi=- 2 \pi T_{rev}/(T_{vib})^3$, i.e. adjusted to match the revival period $T_{rev}$ (see Eq.(\ref{eq:trev})) of the resonant level $v$=98. Indeed, for $\chi$ satisfying $\chi_{v_0}= -2 \pi T_{rev}(v_0)/ [T_{vib}(v_0)]^3$, the amplitude resonantly transferred at $t=t_C (v)$ creates a wavepacket at the outer turning point of each vibrational level $v\approx v_0$, which reaches the inner turning point at time $t = t_P + T_{vib}(v_0)$ independently of $v$.  The evolution of the wavepacket after the pulse has been computed, and we have represented in Fig.\ref{fig:focus} a snapshot of the wavepacket in the excited state at time $t_P + T_{vib}/2$.  
\begin{itemize}
\item An important focussing effect is visible for a negative chirp, the wavepacket presenting a peak in the vicinity of the inner turning point of the $v$=98 level.
\item This effect is reduced when a positive chirp is used, a large part of the population being in the large distance region, due to the late excitation of the upper levels $v$= 98-107, with a much larger vibrational period.
\item When there is no chirp, the population transfer is much smaller,  the factor $\approx$ 3  on the maximum amplitude resulting  into one order of magnitude in the transfer probability ( 4 $\times$10 $^{-5}$ instead of 3.5$\times$ 10 $^{-4}$). Besides the wavepacket is no longer focussed since the levels that have been populated are no longer in the vicinity of $v$=98, but belong to the domain $v >$ 103 of very excited levels; from the behaviour of this wavepacket it is clear, as illustrated in Fig.\ref{fig:focus}(b), that the levels have  a large vibrational period. The distribution of vibrational levels is governed by the spectral width $\hbar \delta \omega$ and by the overlap integral between the initial and final wavefunctions, which is dominant for $v \approx$ 130.
\item Focussing of the wavepacket on the inner classical turning point of the vibrational levels will allow to consider population transfer towards bound levels of the ground triplet state potential curve via a second laser,  in a two colour pump-probe experiment. It should then be possible to  populate efficiently low $v''$ levels in the ground state.
\item For such purpose, the optimization of the pulse can be achieved through analytical formula, considering the revival period, and the scaling laws governing the dynamics of long range molecules \cite{masnou01,stwalley73}.
\end{itemize}

\subsection{Towards new experiments}
A proper estimation of the photoassociation rate requires incoherent average over a thermal distribution of energy normalized initial states. These energy-normalized continuum wavefunctions are deduced from wavefunctions normalized to unit in the box of size $L_R$, accounting for the density of states in this box.  Furthermore, following Ref.\cite{mackholm94}, for low temperature, it is possible to estimate the photoassociation yield by assuming a weak variation of the population as a function of the initial energy. For an assembly of atoms in a volume $V$= 5$\times$ 10$^{-4}$ cm$^3$, with density $N_{at}$=10$^{11}$ cm$^{-3}$, the number of molecules formed in one pulse under the conditions discussed in the present paper is 0.690. With a repetition rate of 10$^8$ Hz, the yield is 6.9$\times$ 10$^7$ molecules per second, which is significant larger that rates obtained with continuous lasers \cite{masnou01}.

\section{Conclusion}
\label{sec:conclu}
The present theoretical paper has investigated the possibility offered by replacing continuous lasers by chirped laser pulses in ultracold photoassociation experiments. We have performed and analyzed numerical calculations of the population transfer due to a chirped pulse, from a continuum state ($T \approx 54 \mu K$) of the ground triplet state Cs$_2$ $a^3\Sigma_u^+$(6s + 6s), to high excited vibrational levels  in the external well of the  0$_g^-$ (6s + 6p$_{3/2}$) potential. The central frequency of the pulse was chosen resonant with the level $v$=98 of the  0$_g^-$ external well, bound by 2.65 \wn. We have used a pulse with gaussian envelope, having a peak intensity $I_L \approx$ 120 kW cm$^{-2}$,  a linear chirp parameter of -4.79 $\times$ 10$^{-3}$ ps$^{-2}$, and temporal width  $\tau_C=34.8$ ps. Due to the linear chirp parameter,  a range  of energy of 2 $\times$ 0.87 \wn \  is  swept by the  laser frequency, corresponding to resonance  with 15 vibrational levels in the vicinity of v$=98$, and referred to as ``resonance window''. The spectral width of the pulse, which remains independent of the chirp, covers a typical range of energies of $\sim$ 0.98 \wn  on  both sides. 

The initial state of two colliding ultracold cesium atoms is represented  by a continuum stationary wavefunction, with a realistic scattering length and nodal structure. This is possible owing to use of a mapped sine grid method recently developed by Willner {\it et al} \cite{willner04}. The time-dependent Schr\"odinger equation is solved through expansion in Chebyschev polynomia \cite{kosloff94}.

 Due to the large value of the scattering length, the initial wavefunction has a large density probability at large distances, resulting into a large Franck-Condon overlap with vibrational levels in the excited state close to the dissociation limit. Numerical calculations show that, during the pulse,  a large amount of population is indeed transferred to levels close to the Cs$_2$ (6s + 6p$_{3/2}$) dissociation limit: however, with the particular choice of the chirp, we have shown that  this population is going back to the ground state after the pulse. In contrast, the 15 levels within the photoassociation window are coherently populated, and part of the population remains after the pulse. The same conclusion is obtained by changing the sign of the chirp. Besides, an interesting result is that, due to the coupling at large distances between the two electronic states,  a strong population transfer to the last vibrational levels of the ground  $a^3\Sigma_u^+$(6s + 6s) is realized, making stable molecules with a rate  as important as for the photoassociation into the excited state. This result should be further explored. \\

The interpretation of the strong population transfer towards the 15 levels in the resonance window, independent of the sign of the chirp, is given within the impulsive limit, in the framework of a two-state adiabatic population inversion model. The numerical results  can be qualitatively and even quantitatively interpreted by this model, defining the conditions for full population transfer within a ``photoassociation window'' covering a range of distances 85 a$_0\le R \le$ 110  a$_0$, and no transfer outside. Therefore the chosen chirped pulse is capable of controlling the population transfer towards the excited state, ensuring that no transfer occurs outside a chosen energy range. We have discussed the general conditions for adiabatic population inversion during a ``time window''  [$t_P-\tau_C, t_P+\tau_C$] centered at the time $t_P$ of the maximum of the gaussian pulse, and corresponding to 98 $\%$ of the energy transferred  by the pulse.   The adiabaticity condition within the ``photoassociation window'' and ``during the time window'' requires that the coupling should be larger than a quantity proportional to the geometric average between the energy range swept by the instantaneous frequency of the chirped pulse and its spectral energy width.  To keep adiabaticity, a compromise between increasing the laser intensity and decreasing the width of the photoassociation window must therefore be found, and it can be discussed within the model we propose. Even though in the present calculations the adiabaticity condition is verified within the photoassociation window, a few non adiabatic effects were identified and interpreted. Further work will estimate more thoroughly the lower limit for intensity.

As  an example of control, the present pulse has been chosen in order to achieve focussing of the excited vibrational wavepacket at the inner turning point of the $v$=98 level,  at a time delay after the pulse maximum equal to half the classical vibrational period $T_{vib}(v=98)$. The relevant chirp parameter was optimised from the revival time.  Such calculations suggest a scheme for a two colour experiment where a second pulse, delayed by $T_{vib}/2$, would transfer photoassociation  population to low $v''$ levels of the ground triplet state, now forming stable molecules in  low vibrational levels.

In a forthcoming paper, we shall give more quantitative results for comparison with experiment by averaging on the initial state velocity distribution to compute the photoassociation rate for a series of pulses with a variable repetition rate, considering  different choices of the final vibrational levels by varying the detuning, and  different laser intensities. We shall propose various schemes to optimize the efficiency of  the population transfer, taking advantage of the conclusion of the present paper where we have shown that it is possible to design chirped laser pulses so that the photoassociation reaction is restricted to a well defined  window. 

From the present calculations  it is already possible to conclude that photoassociation with a chirped laser pulse is a very promising scheme. In the present example where the peak intensity  of the laser was chosen $I_L \approx$ 120 kW cm$^{-2}$, the estimated rate of  $\sim$ 7 $\times$ 10$^7$  molecules per second in typical experimental conditions is definitely larger than what has been computed and measured for cw lasers.  Since the present calculations were performed in a frequency domain close to a photoassociation minimum, much higher rates are to be expected by varying the detuning, and this will be the subject of a forthcoming paper. \\

\underline{\bf Acknowledgements}

This work was peformed in the framework of the Franco-Israeli binational cooperation program Arc en Ciel-Keshet 2003-2004, under contract no. 23, and of the European Research Training Network ``Cold Molecules'', funded by the European Commission under contract HPRN CT 2002 00290. M.V. acknowledges for a three months post-doctoral stay in Orsay funded by this contract.

\bibliographystyle{jcp}
\bibliography{/export/home/fmasnou/latex/Biblio/equipe,/export/home/fmasnou/latex/Biblio/photoass,/export/home/fmasnou/latex/Biblio/coldmol,/export/home/fmasnou/latex/Biblio/bose,/export/home/fmasnou/latex/Biblio/feshbach,/export/home/fmasnou/latex/Biblio/spectroscopy,/export/home/fmasnou/latex/Biblio/potentials,/export/home/fmasnou/latex/Biblio/wavepacket,/export/home/fmasnou/latex/Biblio/scatlength}

\newpage
\newpage
\newpage
\begin{figure}
{\includegraphics*[width=0.80\textwidth,angle=0]{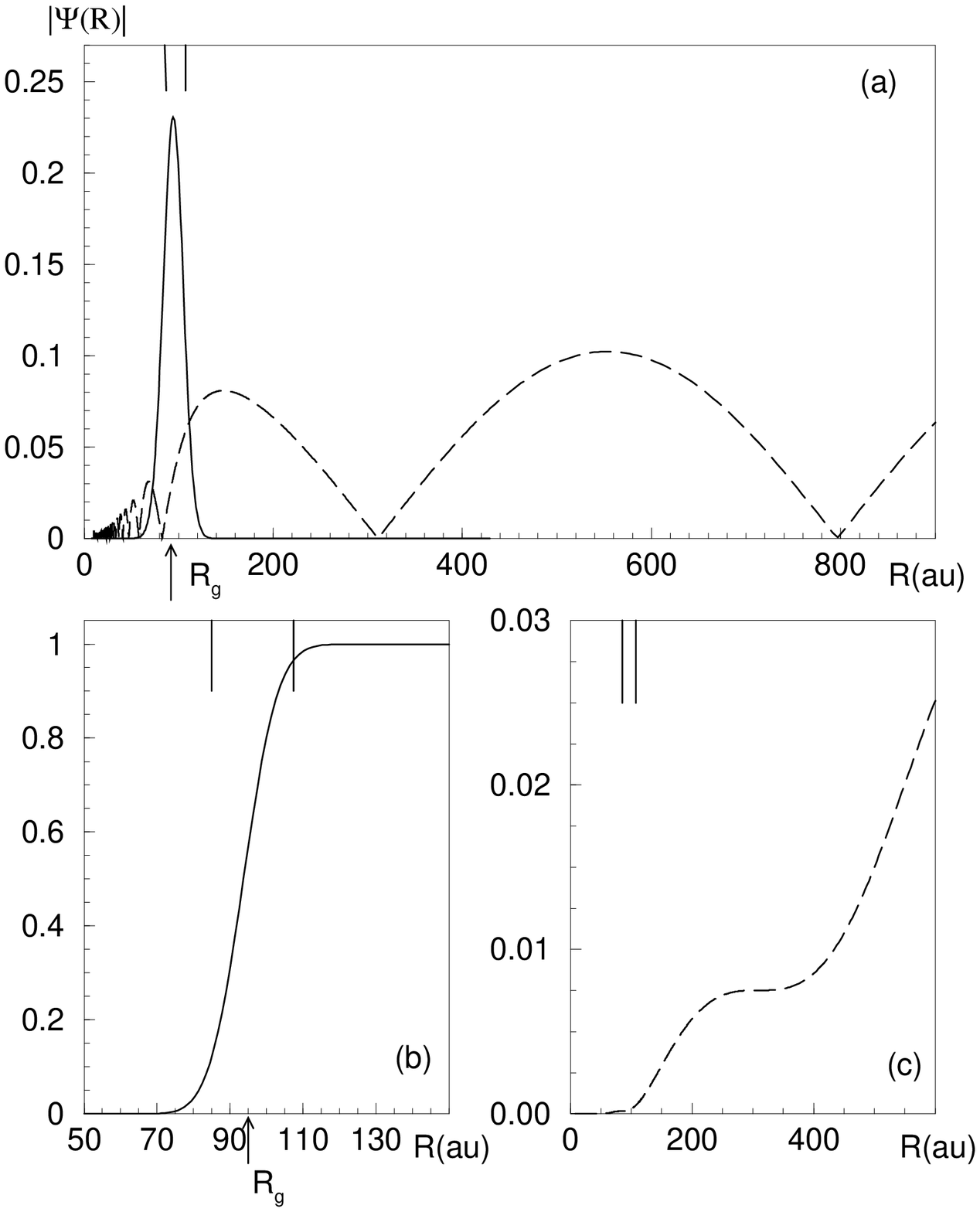}}
\caption {Wavefunction of the system before the pulse; the range of distances $[R_{min}=85 \ a_0, R_{max}=110 \ a_0]$, later on referred to as ``photoassociation window'' is indicated on the upper horizontal scale, the crossing point $R_L$=93.7 $a_0$ on the lower one.\\
 a) Variation of $10 \times |\Psi^{\omega_L}_{g}(R,0)|$ (broken line), the continuum wavefunction   describing the relative motion of two Cs atoms in the potential a$^3\Sigma_u^+$ at energy $E=kT=$ 1.72$\times$ 10$^{-10}$au, $T$ = 54$\mu$K. For visibility, this wavefunction, normalized to unit in a box of size 19 250 $a_0$, is multiplied by 10. An energy-normalized function can be deduced by dividing the unit normalized function by $\sqrt{dE/dn} = \sqrt{ 8.87 \times 10^{-12}}$. Also represented is   a gaussian wavepacket of width $\sigma$ = 15 au, centered at $R_g=95 \ a_0 \approx R_L$. \\
 b) and c) Integrated density of probability $\int^R_0 | \Psi|^2 dR$ for   the localized gaussian wavepacket (b)  and the delocalized continuum wavefunction (c) : note that in the region 85 a$_0\le R \le$ 110  a$_0$, the probability density is only 0.00044 in the last case, while it is 0.86 for the localized wavepacket.}
\label{fig:init}
\end{figure}
\newpage
\vspace{2cm}
\newpage
\newpage

\newpage
\vspace{2cm}
\newpage
\newpage
\vspace{4cm}
\end{document}